\documentclass[twocolumn]{aastex63}
\usepackage[utf8]{inputenc}
\usepackage{graphicx}
\usepackage{natbib}

\shorttitle{Systematic Errors in Faint K2 Light Curves}
\shortauthors{Moreno et al.}

\graphicspath{{./}{FinalFigures/}}

\begin{document}

\title{Properties of a Previously Unidentified Instrumental Signature in \textit{Kepler/K2} that was Confused for AGN Variability}

\received{}
\revised{}
\accepted{}
\submitjournal{}

\author[0000-0001-9134-6522]{Jackeline Moreno}
\affiliation{Department of Physics, Drexel University, 32 S.\ 32nd Street, Philadelphia, PA 19104, USA}

\author{Rachel Buttry}
\affiliation{Department of Physics, Drexel University, 32 S.\ 32nd Street, Philadelphia, PA 19104, USA}
\affiliation{Department of Physics, Carnegie Mellon University, 5000 Forbes Avenue, Pittsburgh, PA 15213, USA}

\author{John O'Brien}
\affiliation{Department of Physics, Drexel University, 32 S.\ 32nd Street, Philadelphia, PA 19104, USA}
\affiliation{Department of Physics and Astronomy, Michigan State University, East Lansing, MI 48824, USA}

\author[0000-0001-7416-9800]{Michael S. Vogeley}
\affiliation{Department of Physics, Drexel University, 32 S.\ 32nd Street, Philadelphia, PA 19104, USA}

\author[0000-0002-1061-1804]{Gordon T. Richards}
\affiliation{Department of Physics, Drexel University, 32 S.\ 32nd Street, Philadelphia, PA 19104, USA}

\author[0000-0001-5785-7038]{Krista Lynne Smith}
\affil{Department of Physics, Southern Methodist University, Dallas, TX 75205, USA}

\begin{abstract}

The {\em Kepler} satellite potentially provides the highest precision photometry of active galactic nuclei (AGN) available to investigate short-timescale optical variability.  We targeted quasars from the Sloan Digital Sky Survey that 
lie in the fields of view of the {\em Kepler}/K2 campaigns.
Based on those observations, we report the discovery and properties of a previously unidentified instrumental signature in K2.
Systematic errors in K2, beyond those due to the motion of the detector, plague our AGN and other faint-target, guest-observer science proposals. Weakly illuminated pixels are dominated by low frequency trends that are both non-astrophysical and correlated from object to object.
A critical clue to understanding this instrumental noise is that
different targets observed in the same channels of
Campaign 8 (rear facing) and Campaign 16 (forward facing) had nearly identical light curves after time reversal of one of the campaigns.  
This observation strongly suggests that the underlying problem relates to the relative Sun-spacecraft-field orientation, which was approximately the same on day 1 of Campaign 8 as the last day of Campaign 16.  Furthermore, we measure that the instrumental signature lags in time as a function of radius from the center of the detector, crossing channel boundaries.  Systematics documented in this investigation are unlikely to be due to Moir\'{e} noise, rolling band, or pointing jitter. Instead this work strongly suggests temperature-dependent focus changes that are further subject to channel variations. Further characterization of this signature 
is crucial for rehabilitating K2 data for use in investigations of AGN light curves.

\end{abstract}

\section{Introduction}
The {\em Kepler} satellite was originally conceived as an  exoplanet finding mission.  Its rich archival data (observing 2009-2018) has revolutionized planet discovery and characterization.  {\em Kepler}'s second mission, K2, has also impacted many other research fields including astrosiemsology, stellar activity, microlensing, white dwarfs, supernovae, active galactic nuclei (AGN), and more, together comprising $\sim60\%$\footnote{https://keplerscience.arc.nasa.gov/publications.html\#breakdown-by-subject}
of all publications using K2 data thus far. However, poorly understood instrumental systematics posed a significant impediment to an efficient exploitation of K2's thousands of light curves.  Part of the problem is that noise removal is optimized differently depending on the science objective.

Noise removal takes on one definition for planet transit detection and another for characterization of a faint ``unknown" signal, as is the case for AGN in the short timescale regime ($<100$ days; optical wavelengths). A planet transit manifests as a strictly periodic signal, characterized by a peak at the orbital frequency in the power spectral density (PSD) of a light curve. In contrast, AGN exhibit intrinsically \textit{noisy} fluctuations that are best characterized by the entire shape of the empirical power spectrum.  The problem of cleaning data of instrumental noise at nearly all observed frequencies is critical to enabling science that relies on the entire PSD, which is the case for aperiodic or stochastic variability.

K2 data provides the most evenly sampled high-cadence (30 minutes) observations for a large sample of AGN in the optical (over 4000 targets).  If these data are properly mitigated for systematics, the observed sample may be studied jointly with data from the Zwicky Transit Factory \citep{Bellm2019} and the Deep Drilling Fields of the Vera Rubin Observatory \citep{Ivezic2008} to probe supermassive black hole (SMBH) physics on timescales of days to months.  The physics in this regime is extremely valuable to constraining the corona-accretion disk relationship \citep[e.g.,][]{GP19} across the broad spectral taxonomy of AGN (blazars, BL LAC, Seyfert I, Seyfert II, radio quiet quasars, radio-loud quasars, weak-lined quasars, broad-absorption line quasars, and more).

Characterizing the short-timescale variability properties of AGN offers the promise of yielding deeper insights into the accretion physics of SMBHs and can potentially constrain physical scales. 
Most importantly, {\em Kepler}/K2's extremely regular 30-minute cadence and sensitivity (to quasars) will remain unmatched for the foreseeable decade.  Between \citealt{Mushotzky2011},\citealt{Wehrle2013}, \citealt{Revalski2014}, \citealt{Edelson2014}, \citealt{EShaya2015},  \citealt{Chen2015}, \citealt{Vish2015}, \citealt{Dobrotka2017}, and \citealt{Smith2018}, twenty-one (21) Kepler AGN have already revolutionized our understanding of AGN variability, while $\sim$ 4000 K2 AGN remain un- or under-studied.

Because of the failure of the second of four reaction wheels in 2013, causing {\em Kepler} to be re-purposed as the ``K2" mission, the telescope was constrained to observe along the orbital plane while steadying pointings with solar radiation pressure and periodic thruster firings. Several re-processing pipelines were developed to correct pointing errors known as arc-drift (e.g., \citealt{VJ2014}; \citealt{Armstrong2015}; \citealt{Crossfield2015};
\citealt{Foreman-Mackey2015}; \citealt{Huang2015}; \citealt{Lund2015}; \citealt{Aigrain2015}, \citealt{Libralato2015}).   Custom reprocessing software and recommendations were also published in the literature \citep{Kinemuchi2012}. Although many of these pipelines were optimized specifically for exoplanet transit detection, the majority were highly successful at producing arc-drift-free light curves and enabling science for many types of highly variable astrophysical signals. Many of these re-extracted light curves are publicly available on \textit{K2-MAST}\footnote{https://archive.stsci.edu/missions-and-data/k2}\footnote{https://keplerscience.arc.nasa.gov/new-keplerk2-high-level-science-products-available.html}.

Large-sample science that may be possible with data in multiple K2 campaigns requires a streamlined approach that manual custom re-processing cannot scale to. However, no single prescription to date rehabilitates K2 data for the study of all types of astrophysical variability. Correlated {\em Kepler}/K2 noise systematics may arise from a combination of error sources such as rolling band, Moir\'{e} noise, faint blended background sources \citep{KeplerHandbook}, and point spread function (PSF) changes \citep{Libralato2015}.
To characterize systematics, we primarily take advantage of correlations across multiple sources on the same channel. Using all pixels downlinked by K2 (per campaign), we detect features that exhibit dependencies on campaign, channel, magnitude, and distance from the boresight that cause hundreds of targets to be correlated with each other. Low-frequency and red-noise systematics, which are not mitigated by arc-drift corrections (\citealt{VJ2014}, \citealt{Lugar2018}), especially plague AGN and other faint targets \citep{jack2018}.

Unfortunately, there is no pipeline that sufficiently cleans K2 data for large-sample AGN variability studies.  In this work, we discuss possible ways to identify systematics in K2 light curves that have been confused for astrophysical variability in published literature (e.g., \citealt{Aranzana2018} for K2 AGN and \citealt{Dobrotka2017} for Kepler AGN followed by a retraction in \citealt{Dobrotka2019}).  
On particular channels, systematics may dominate even at the optimal Kepler magnitude (12th magnitude stars). Finally, we review available light-curve-processing software that may provide suitable starting points for rehabilitating K2 data and we provide recommendations for continued development of reprocessing software that may potentially produce a fully usable light curve PSDs for scientific investigation. 

This paper is organized as follows. Section~\ref{sec:data} describes K2 AGN light curves that are available in Campaigns 0-19 with an emphasis on targets that overlap with Sloan Digital Sky Survey (SDSS; \citealt{York2000}) fields.  Section~\ref{sec:cbvs} discusses known systematics in the first phase of {\em Kepler} that were corrected with cotrending basis vectors (CBVs); Section~\ref{sec:correlatedAGN} presents our investigation of remaining systematics in K2 and discusses possible approaches for mitigating errors more effectively than CBVs.  Section~\ref{sec:software} demonstrates the performance of popular reprocessing methods such as the self-flat-fielding prescription of \citet{VJ2014} and the {\tt Everest 2.0} pipeline of \citet{Lugar2018}.   
Finally Sections~\ref{sec:why} and \ref{sec:summary} summarize our AGN science motivations to rehabilitate K2 data and the most important findings in this work.

\section{Data}
\label{sec:data}

The {\em Kepler} telescope focused on a fixed 116 square degree field of view (FOV) between Cygnus and Lyra for 4 years prior to K2.
The focal plane array consisted of forty-two $2200\times 1024$ pixel CCDs\footnote{For focal plane labelling and signal map per CCD, see Figures~24 and 25 in  https://archive.stsci.edu/kepler/manuals/KSCI-19033-001.pdf}. Each CCD had a half-maximum bandpass of 435 nm to 845 nm and was read out through two channels (84 CCD channels also referred to as ``module outputs").
The long-cadence data consisted of 270 frame (6.02 second exposures) integrations for an epoch every $\sim$ 30 minutes \citep{KeplerHandbook}. 
The first Kepler mission observed continuously for a period of $\sim$90 days (a time frame referred to as a quarter), then downlinked to Earth, followed by a 90 degree spacecraft rotation before returning to observing mode.  A typical quarter contained $\approx4000$ epochs known as cadences. 
However, not all 96 million pixels per cadence were downlinked due to limited bandwidth. A specific image cutout, referred to as a postage stamp or mask, of fixed size for the duration of the quarter or campaign was saved for storage. 
The longest AGN light curve available, Zwicky 229-015, contains 14 quarters of data (3.9 yrs; e.g., \citealt{Vish2015}).

A systematic effort was undertaken by several groups to identify bright AGN in the  \textit{Kepler} Field of View (FOV). By cross-matching observations from ROSAT, 2MASS, and WISE, \citet{Edelson2012} identified AGN in the Kepler FOV. Radio-loud AGN were identified in the Kepler FOV using the VLBA \citep{Wehrle2013}. Further efforts to identify AGN candidates brought the total number up to 87 (Kepler GO40041 Edelson). Twenty of these AGN are spectroscopically confirmed (7 cataloged in \citet{VCV2010} and 13 by \citealt{Edelson2012}). The KSwAGS X-ray survey of the Kepler field conducted with SWIFT discovered several candidate AGN, 13 of which have been spectroscopically verified, and many X-ray bright stellar sources \citep{KSmith2015}.
Data for all observations acquired during quarters 0 through 17 are publicly available on the Kepler MAST archive (Data Release 25).


\begin{figure}[th!]
    \epsscale{1.05}
    \plotone{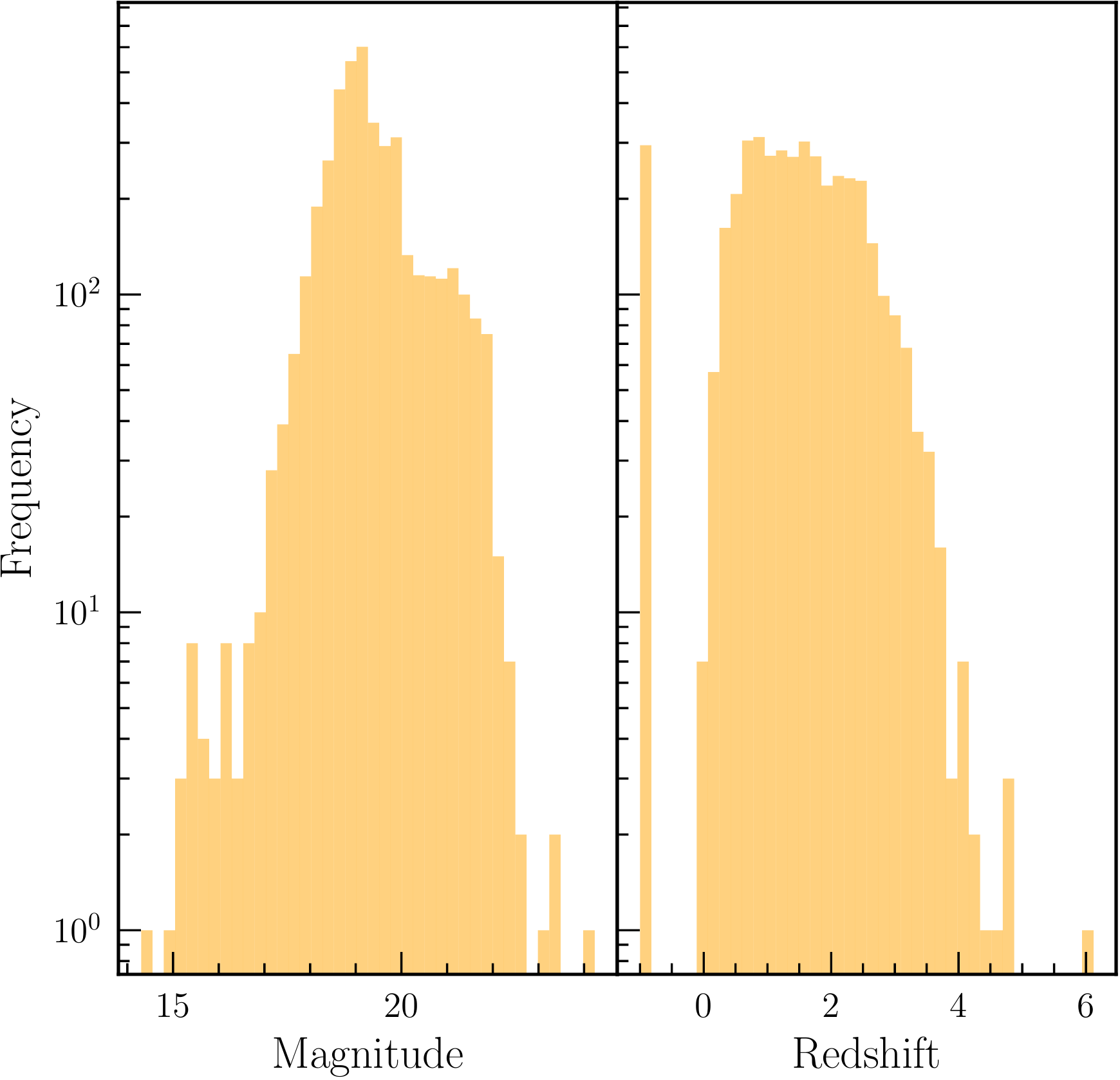}
    \caption{Left: Apparent magnitude distribution of AGNs from our K2 proposal that were observed.  Right: Redshift distribution of AGNs from our K2 proposal that were observed. Quasar candidates with no available spectroscopic redshifts have a value of -1. The confusion limit of Kepler is reported at about 20th magnitude.}
    \label{fig:fig1}
\end{figure}

\begin{table}\caption{AGN Targets by Campaign}
\begin{center}
\label{tab:AGNtargs}
\begin{tabular}{lccc}
\hline
\hline
Campaign & Obs. Targs & zmax & imagmax\\
\hline
8 & 444 & $\sim$4 & $\sim$19.5\\
10 & 426 & $\sim$4 & $\sim$19.5\\
12 & 337 & $\sim$4 & $\sim$20.5\\
14 & 622 & $\sim$4 & $\sim$20.0\\
16 & 1047 & $\sim$4 & $\sim$20.0\\
18 & 123 & $\sim$3 & $\sim$19.0\\
19 & 1169 & $\sim$6 & $\sim$22.0\\
\hline
\end{tabular}
\end{center}
\end{table}

Pointing jitter was a minor source of error during the first \textit{Kepler} mission, causing a maximal 0$\farcs$55 drift of the target centroids across pixels (3$\farcs$92 in length).  
After the failure of the second of four reaction wheels in May 2013, engineers (Ball Aerospace; \citealt{McCalmont2015})  devised a way to stabilize pointings. Spacecraft roll error resulted from residual radiation pressure torques, known in the literature as arc-drift \citep{Cleve2016}. The \textit{Kepler} pipeline detrending module (PDC; \citealt{Smith2012}) was not updated to correct for arc-drift and the first K2-handbook detailing data quality was published in 2017 (and updated in 2018; \citealt{Cleve2017}). 

Kepler's second mission, K2, deployed an observing strategy consisting of 19 subsequent observing fields along the ecliptic with an $\approx80$ day duration. Each of these observation periods is referred to as a campaign.  
Campaign 0 provided a commissioning data set used to evaluate the pointing jitter and necessary adjustments.  It pointed at the Galactic `Anti-Center' and observed 113 AGN targets of various types ranging from relatively ordinary Seyferts 1 to highly variable blazars. The Campaign 1 FOV pointed at a region of the North Galactic Cap.  The total number of extra-Galactic targets observed was $\sim$ 550 with $\sim$ 300 thought to harbor AGN. Campaigns 2 $\&$ 3 observed another 266 AGN of various sub-types including bright quasars and more blazars.  

Our group proposed additional observations of known quasars beginning with Campaign 8.
These quasar targets were selected from the SDSS-I/II \citep{srh+10} and SDSS-III
quasar catalogs \citep{DR12Q}.  Robust photometric quasar candidates
(with positions accurate to $\sim$50 mas) are also drawn
from \citet{rng+04}, \citet{bhh+11}, \citet{Richards15}, and
\citet{Peters15}, where the Peters et al.\ catalog specifically
targeted quasar candidates due to their variability.
Thus, our targets are a heterogeneous K2 sample of known quasars and
high-confidence photometric quasar candidates (most with robust photometric
redshifts).  In short, we proposed to target all SDSS quasars (and many quasar candidates) that overlapped with the K2 field of view in each campaign starting with Campaign 8.  The focus was on objects in SDSS Stripe 82 where there already exist $\sim$ 100 epochs of photometry \citep{annis+14} that can be coupled with K2 photometry to increase the time baseline for investigations of quasar variability.  
Figure~\ref{fig:fig1}
illustrates the magnitude and redshift distribution of SDSS quasars targeted by our group in Campaigns 8, 10, 12, 14, 16, 18, and 19.  These targets include objects that are fainter than the typical range of magnitudes included in the standard pipeline processing \citep[e.g.,][]{Lugar2018}.  Table~\ref{tab:AGNtargs} gives the number of observed targets and approximate maximum redshift and faint limit ($i$ band) for each campaign.  Note that many more targets were proposed than were accepted and the process for sub-selecting the observed targets was not uniform (sometimes favoring bright targets and known quasars, but other times to reduce high spatial density such as in the Stripe 82 area).

\section{Kepler/K2 Cotrending Basis Vectors}
\label{sec:cbvs}

Low frequency instrumental systematics in the first Kepler mission could be removed (to an extent) with cotrending basis vectors (CBVs) generated by the data calibration module, PDC (\citealt{Stumpe2012part1}; \citealt{Smith2012}).  CBVs were calculated for each observing quarter and per channel with a principal component analysis (PCA) of an ensemble of ``quiet" stars.  PDC-MAP (maximum a-posterior) supplied CBV-corrected light curves with weights derived from the most correlated neighboring sources to individual targets for a predetermined aperture (using a minimum of 5 vectors).

CBV subtraction was, however, problematic for preserving intrinsic astrophysical variability.  Higher-order vectors (3rd or higher) added variance and high frequency features \citep{Smith2012}. Any coincidental correlations between the intrinsic signal and CBVs was projected onto the model used for correction. CBVs do not mitigate spatially dependent systematics such as (boresight-dependent) focus changes, rolling band, or Moir\'{e} noise.

The challenge of error mitigation in the first phase of {\em Kepler} was typically approached with custom manual re-processing or calibration with ground-based data when available (\citealt{Smith2018} and \citealt{Vish2015}, respectively).  For AGN, custom-aperture or PSF photometry was particularly important in order to minimize noise from background sources in crowded regions and from extended host galaxies. Custom-aperture optimization and CBV subtraction was possible with PyKE \citep{Still2012} or similar software but without the advantage of simultaneously fitting neighboring sources as in PDC-MAP that could mitigate issues of over-fitting CBV components.

The compromise of minimizing known and new instrumental noise while preserving the astrophysical signal of interest carried over into {\em Kepler}'s second mission, K2.  
While correcting complex arc-drift systematics became the calibration priority, few publications focused on methodologies for removing other types of instrumental noise or documenting evidence of {\em Kepler}'s natural degradation. 

CBVs are available for K2 in the {\tt EVEREST} 2.0 package \citep{Lugar2018} calculated with the {\tt SysRem} algorithm \citep{Tamuz2005} from an arc-drift-corrected light curve ensemble. While the first- and second-order CBVs do an excellent job at removing the lowest frequency trends, they cannot account for some of the more troubling and recognizable instrumental signatures that we identify in our AGN and other variable star samples. For both \textit{Kepler} and K2 data, CBVs removed the lowest frequency systematics very well, but residual instrumental trends are at best mixed with variance by the PCA-detrending approach.  Theoretically, PCA cannot account for any local spatial dependence within a CCD (channel) or for time-dependent systematics across the focal plane such as temperature-dependent effects. There is a critical need for other calibration algorithms in order to fully exploit the rich K2 archive.  


\section{Correlated K2-SSF AGN}
\label{sec:correlatedAGN}
\begin{figure*}[ht!]
    \plotone{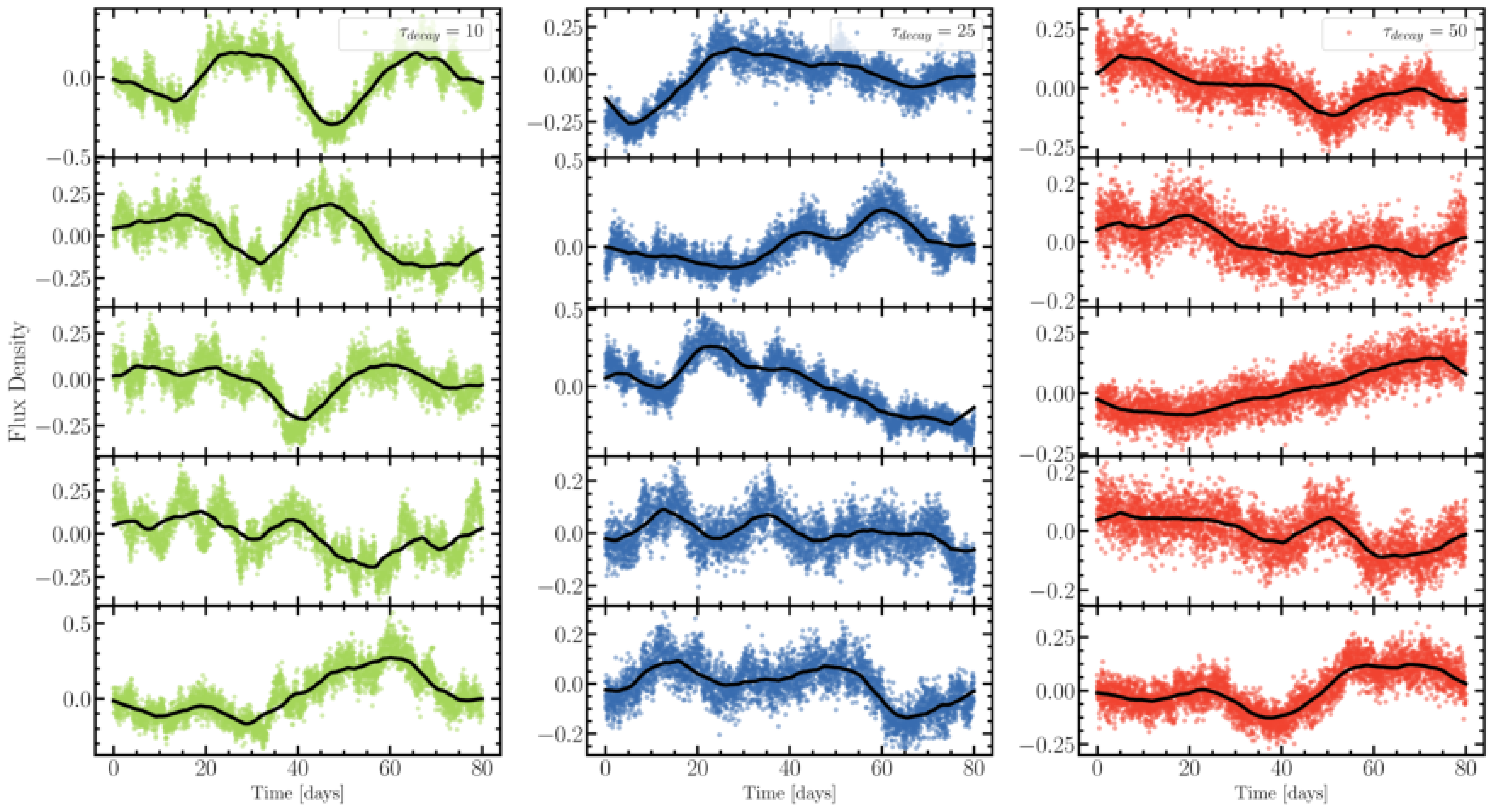}
    \caption{Expected AGN Variability (simulated).  Columns (left to right) show example light curves with increasing characteristic intrinsic timescales $\tau_{decay} = 10, 25, {\rm and } 50$ days. These simulated lightcurves include a constant photometric noise of $5\%$ and constant variability amplitude. The black curve outlines a smoothed trend through each lightcurve to show that relatively low freqency trends do not correlate even within columns for ``objects" having the same characteristic timescales.  We do not expect AGN to vary synchronously, except if that trend is due to correlated noise in the CCD.
    }
    \label{fig:fig2}
\end{figure*}

\begin{figure*}
    \plotone{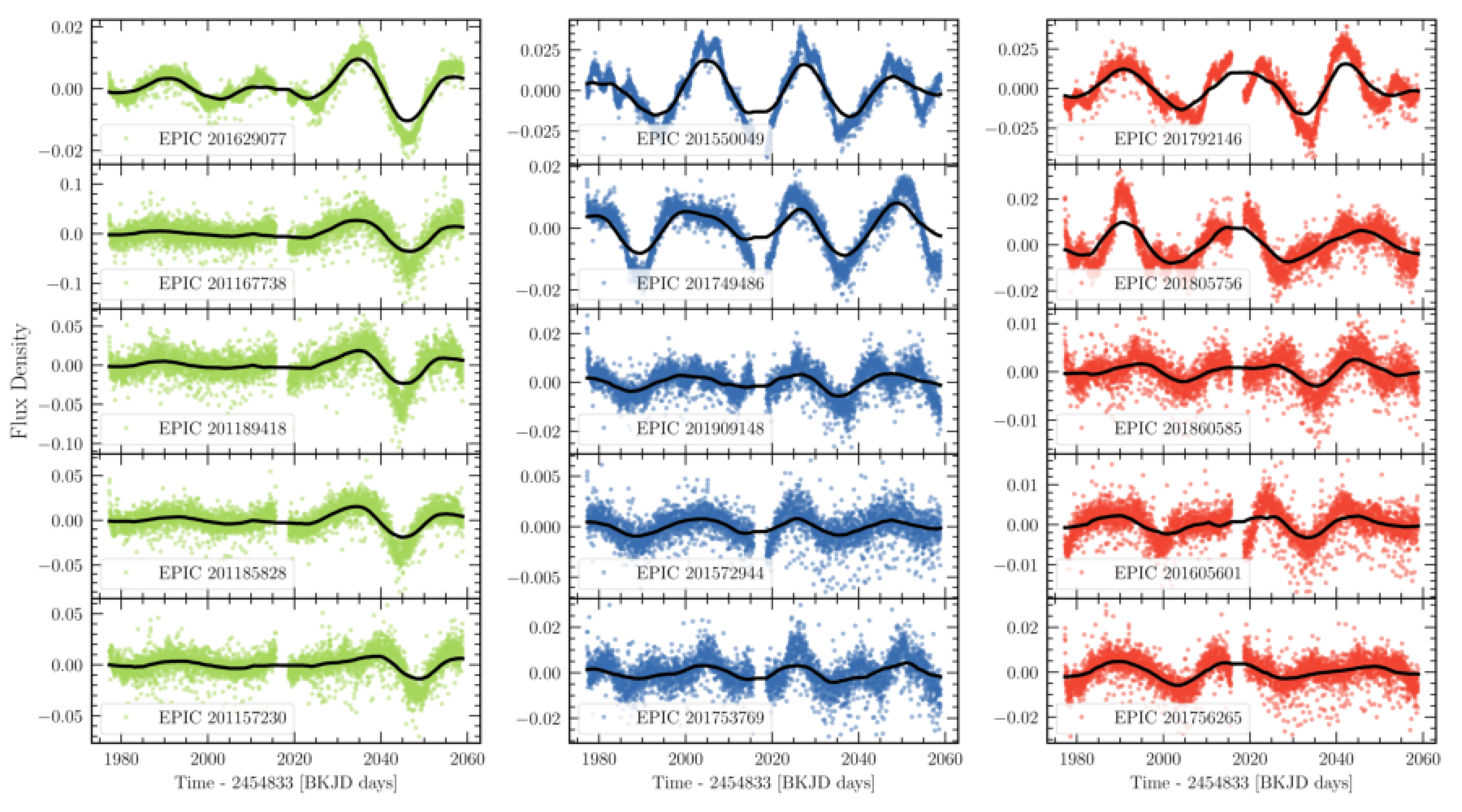}
    \caption{Clustering of full postage-stamp AGN light curves from K2 Campaign 1, referenced in \citet{Aranzana2018}. The Vanderburg and Johnson light curves retain similar patterns that can be detected by the same clustering algorithm. The three different colors highlight just a few of many ``types" of systematic trends that cause independent AGN to be correlated with each other.}
    \label{fig:fig3}
\end{figure*}

The first evidence that we encountered for the presence of residual systematics in K2 data after arc-drift removal came from visually identifying strong correlations between independent target light curves.  Inspection of AGN light curves from K2 indicated that certain patterns were commonly seen, despite the fact that all AGN light curves are expected to be unique.  Further investigation shows that those common patterns reveal strong systematic effects in the data. 

To illustrate the problem, here we compare some simulated light curves that, by design, lack features correlated in time, with some K2 light curves that clearly exhibit correlated features. In Figure~\ref{fig:fig2}, we illustrate the AGN variability that we expected to observe with K2.  Following \citet{Moreno2019}, each column is generated with a simple stochastic model (damped random walk, hereafter DRW) corresponding to characteristic timescales $\tau_{decay}$ $= 10, 25,$ and  $50$ days, respectively. Although the light curves in each column have the same underlying parameters, each is unique and there are not obvious correlations between the different light curves in terms of the locations of the peaks and valleys.

In contrast, in Figure~\ref{fig:fig3} we plot light curves identified using an agglomerative clustering analysis that detects high correlation in the K2 AGN sample used for scientific analysis by \citet{Aranzana2018}.  A visualization of the raw and smooth AGN light curves reveals trends that exhibit multiple peaks and toughs in phase across several sources.   Such patterns are not expected between unique AGNs and indicate that systematics are completely dominating these light curves.  Simply put, any matching variability patterns have nothing to do with the astrophysical targets.

The groups represented in  Figure~\ref{fig:fig3} show light curves from clusters with the strongest covariance.  In the right-most panel, we plot cluster members that exhibit a mixture of astrophysical signal and instrumental trends. Although this clustering analysis was executed on full postage-stamp raw light curves, rerunning the analysis with \textit{K2-SFF} (\citealt{VJ2014}) from \textit{MAST} also reveals the same correlated behavior that we attribute to instrumental systematics. Thus, we find strong evidence that faint-object K2 light curves show patterns that are instrumental, not astrophysical, in origin.  These systematics are not removed by any currently used analysis pipeline.  The goal of this work is to identify degrees of freedom that may prove useful for characterizing and removing this time dependent correlated noise signature from faint targets like AGN.

\subsection{Channel Dependence and Spacecraft Orientation}
\label{Sec:rolling}

\begin{figure*}
    \plotone{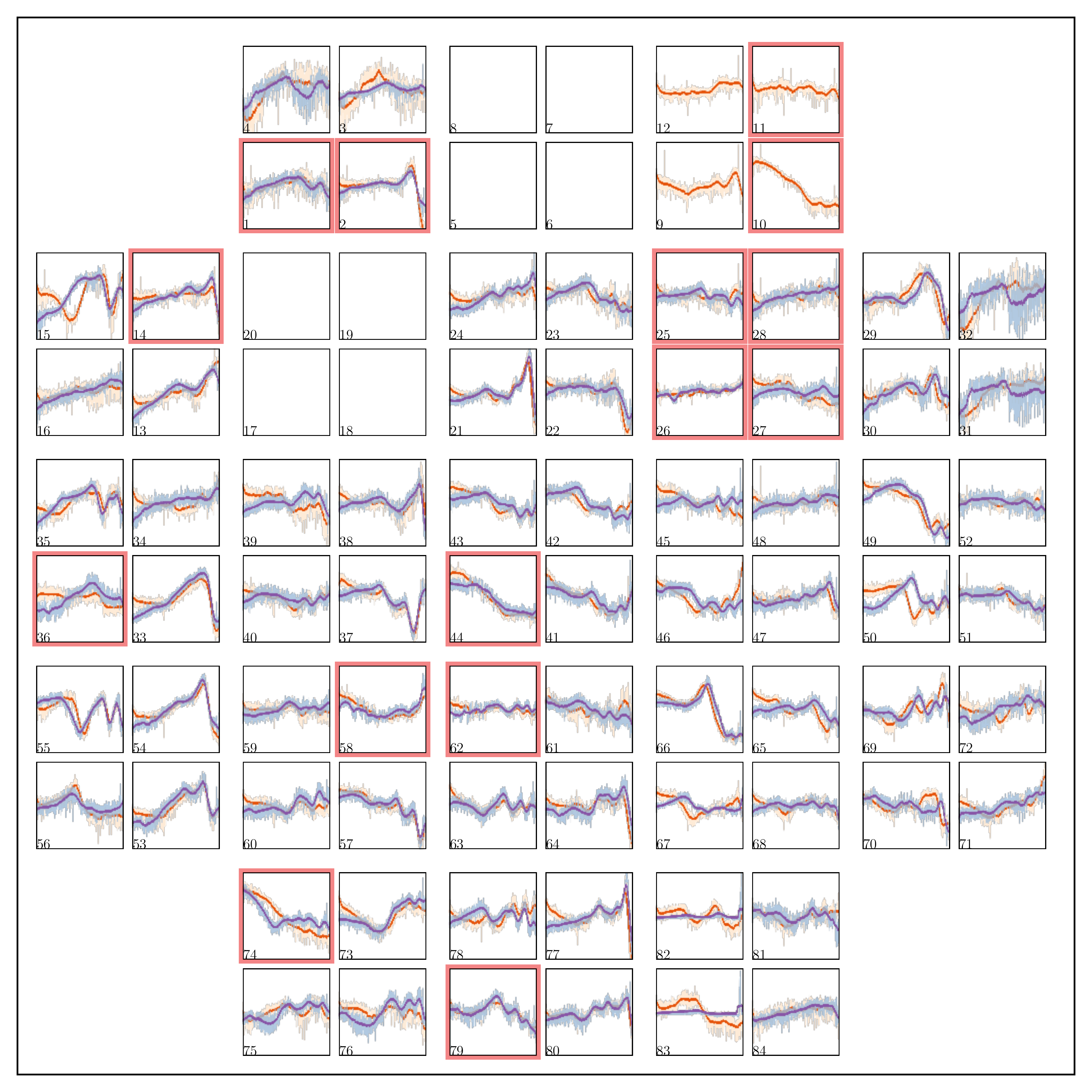}
    \caption{Median channel light curves for K2 Campaigns 8 and 16. The median light curves are calculated by computing the total-postage-stamp light curves for all objects between 13th and 20th magnitude that are observed on a given channel, then taking the median of those light curves. Using all of the pixels in the postage stamp for each object results in individual light curves that are dominated by the systematic noise found in the background pixels. Taking the median of these light curves will result in a light curve that maintains any pattern that is similar across the individual light curves. The median light curves for campaign 8 (orange) and campaign 16 (purple) are plotted on their respective channels. The channels outlined in red are labeled as being affected the worst by rolling band according to the Campaign 0 release notes. The campaign 16 median light curves are time-reversed to illustrate the mirroring of the systematics seen in the two campaigns.}
    \label{fig:fig4}
\end{figure*}

As a next step in our analysis, we test for channel dependence of the instrumental systematics in K2 light curves.  In
Figure~\ref{fig:fig4}, we show median light curves (normalized flux) per channel for Campaign 8 (orange) and time-reversed for Campaign 16 (purple). 
Features resembling an ``M" or ``W" (or other complex patterns) that are synchronized in time reveal channel-dependent systematics that dominate over the astrophysical variability of the sources.

For a given channel in Campaign 8 or 16, we calculate the curves plotted in Figure~\ref{fig:fig4} as follows:
\begin{itemize}
\item [1] We calculate all the light curves for objects within the 13--20 magnitude range on that channel, using all of the pixels in each object's postage stamp (also interpolating any missing/spurious cadences).
\item [2] We normalize the individual light curves by subtracting the mean flux of the object and dividing by the standard deviation. This normalization was done to ensure that the median light curves would not be disproportionately affected by brighter objects.

\item [3] For each channel, we take the median value across all the light curve ensembles, evaluated at each cadence.
\end{itemize}
Using the total postage stamp for a large range of source magnitudes ensures that the systematics from the background pixels will dominate at each cadence.

Note the strong correlations after time-reversal between targets that are unique to the fields corresponding to Campaign 8 (rear facing) and Campaign 16 (forward facing).  This observation is perhaps the most crucial clue to understanding the correlated noise shown in Figure~\ref{fig:fig3}.  It strongly suggests that the underlying problem relates to the relative Sun-spacecraft-field orientation which was approximately the same on day 1 of Campaign 8 and the last day of Campaign 16 and vice versa.  It further means that these systematics are unlikely to be due to Moir\'{e}, rolling band, dust particles, etc.---unless those are somehow dependent on Sun-spacecraft-field orientation and are largely repeatable.  Instead it is strongly suggestive of temperature-dependent focus changes (that are further subject to differences between the channels). This pattern may explain the observation by \citet{Armstrong2015} and \citet{Aigrain2015} regarding the need to split campaigns into separate time segments on either side of when the spacecraft was perpendicular to the sun (see \citealt{Howell+14} Figure~2 for a visualization of the geometry).

\subsection{Rolling band}

According to the Kepler Handbook, rolling band can be a significant source of additive alias noise that depends on channel, local detector electronics (LDE) board temperature, amplifier temperature, and excitation by bright sources.  The rolling band may add as much as 20 flux counts per pixel\footnote{https://keplerscience.arc.nasa.gov/new-in-lightkurve-identifying-time-variable-background-noise.html} and has been measured to propagate across pixels on the timescale of hours \citep{Cleve2016} which was critically important to Kepler's original mission. It is also possible that increases in flux due to rolling band may occur on timescales of days to weeks which are much more detrimental to the K2 mission, heavily obscuring the signal of faint or extragalactic targets. In the case of AGN, 20 counts of noise per pixel would completely obscure the astrophysical signal if all of the source variability is in the range of one hundred flux counts. Examples of such ranges exist among the 21 confirmed AGN of the original Kepler mission (see Figure 12 in \citealt{Krista2018b}). Any additive instrumental signature that modulates the flux of distant AGN over time by a few tens of counts per pixel could entirely dominate the shape of the K2 light curve.  

The channels outlined in red in Figure~\ref{fig:fig4}
are flagged for worst rolling band (Campaign 0 release notes; \citealt{Cleve2017}); however, these channels are not necessarily the most dominated by the type of systematics that we aim to characterize. Since rolling band flags were generated at commissioning, it is possible that noise characterization was only exploited at timescales most likely to compromise the original mission (detection of planet transits) and that low frequency modes of the rolling band are poorly characterized across each CCD. Additionally, channels may not have degraded uniformly and temperature changes may follow very different trends in the K2 mission than they did during pre-launch commissioning. Thus, channels that are unflagged may still be strongly affected by rolling band.

Regardless of timescale, hours or weeks, much of the rolling band should in theory (and as per handbook recommendations) be mitigated by careful background subtraction.
Rolling band is characteristically responsible for a flux increase that propagates across CCD rows, but the systematics seen in Figures~\ref{fig:fig3} and \ref{fig:fig4} remain even after background subtraction ({\tt Everest} light curves, see \citealt{Luger2016} section 3.1 for details of photometry). The low frequency patterns seen in Figures~\ref{fig:fig3} and \ref{fig:fig4} are either overlooked modes of rolling band that are not sufficiently mitigated by a regular background subtraction or they are an altogether different type of noise that turned up post-launch, especially if they cross channel boundaries.

\subsection{Spatial Dependence}

\begin{figure}
    \epsscale{0.9}
    \plotone{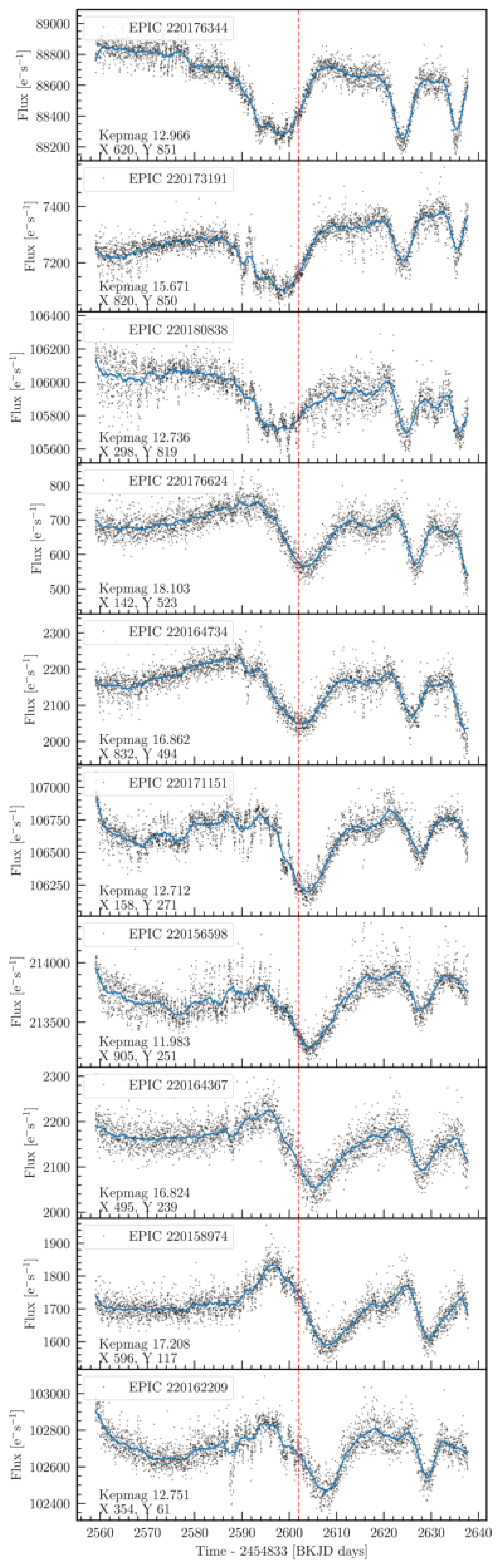}
    \caption{Campaign 8 Channel 55 objects ordered by their Y-position on the CCD. `X' is the pixel column and `Y' is the pixel row. The red line marks the position of the central dip in the light curve of the 4th object. There is a position- (but not magnitude-) dependent phase shift in the systematics.}
    \label{fig:fig5}
\end{figure}

There are a number of arguments against rolling band being the main cause of systematic noise in our AGN sample.  The main argument is that we find a spatial dependence relative to the boresight rather than along the read-out direction of the CCDs. We also note a time dependence of that spatial dependence, which we focus on in this section.  In Figure~\ref{fig:fig5}, we gather a set of noise-dominated light curves from channel 55.  The light curves are ordered by decreasing pixel row index (Y).  The sequence of dip features (starting with three such dips in the top panel) shift by a few days with decreasing pixel row index (ending with only two dip features in the bottom panel example).
Such a dependence may appear to be consistent with ``rolling band" despite channel 55 not being identified as having high rolling band noise. However, the timescale of the shift is days not hours (although there exist examples of what is thought to be rolling band on such timescales), and this shifting sequence of dips in the light curves is seen in reverse in Campaign 16.  Furthermore, we find evidence (below) that the pattern crosses channel boundaries and is related to boresight distance rather than local CCD direction.

\begin{figure}
    \centering
    \epsscale{1.2}
    \plotone{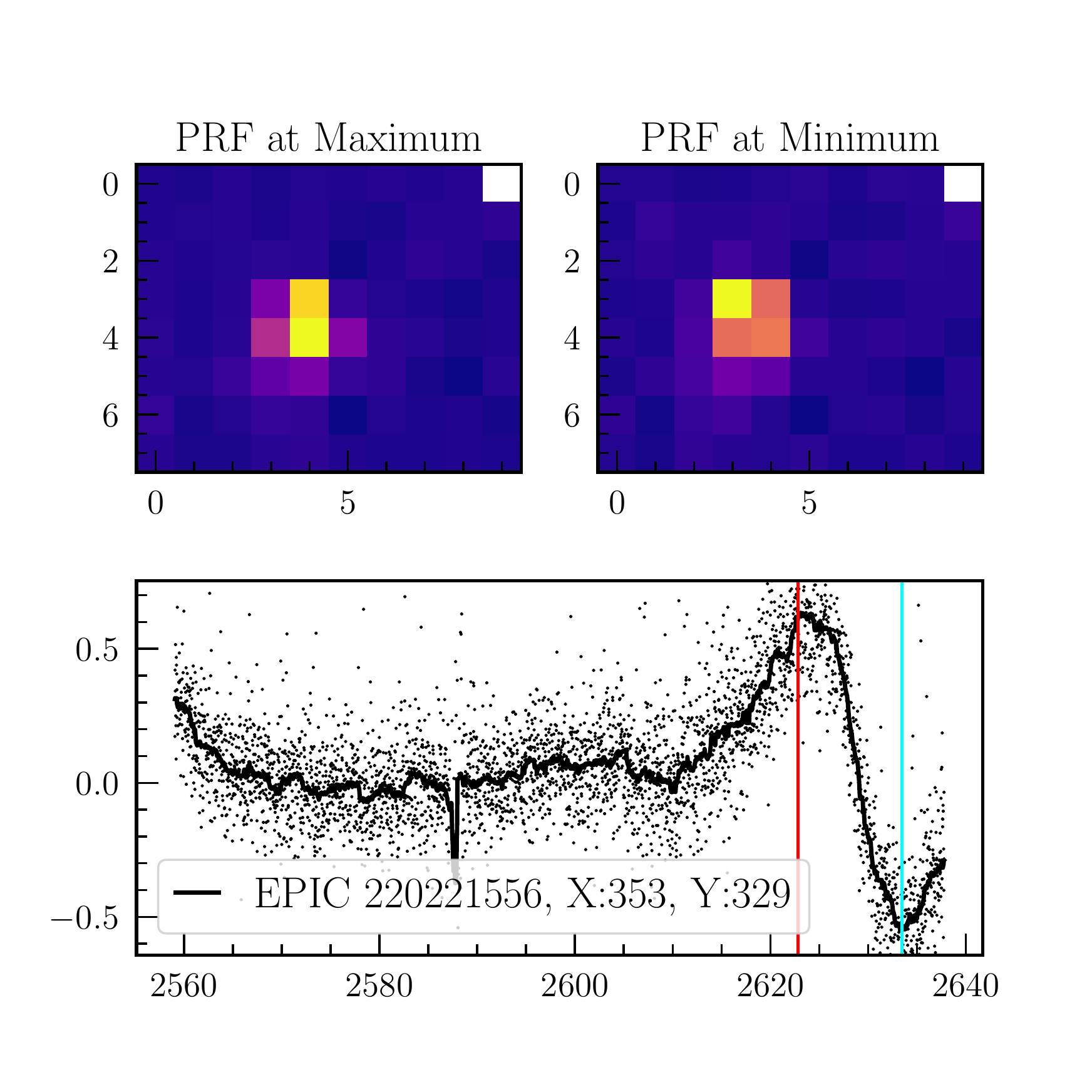}
    \epsscale{1}
    \plotone{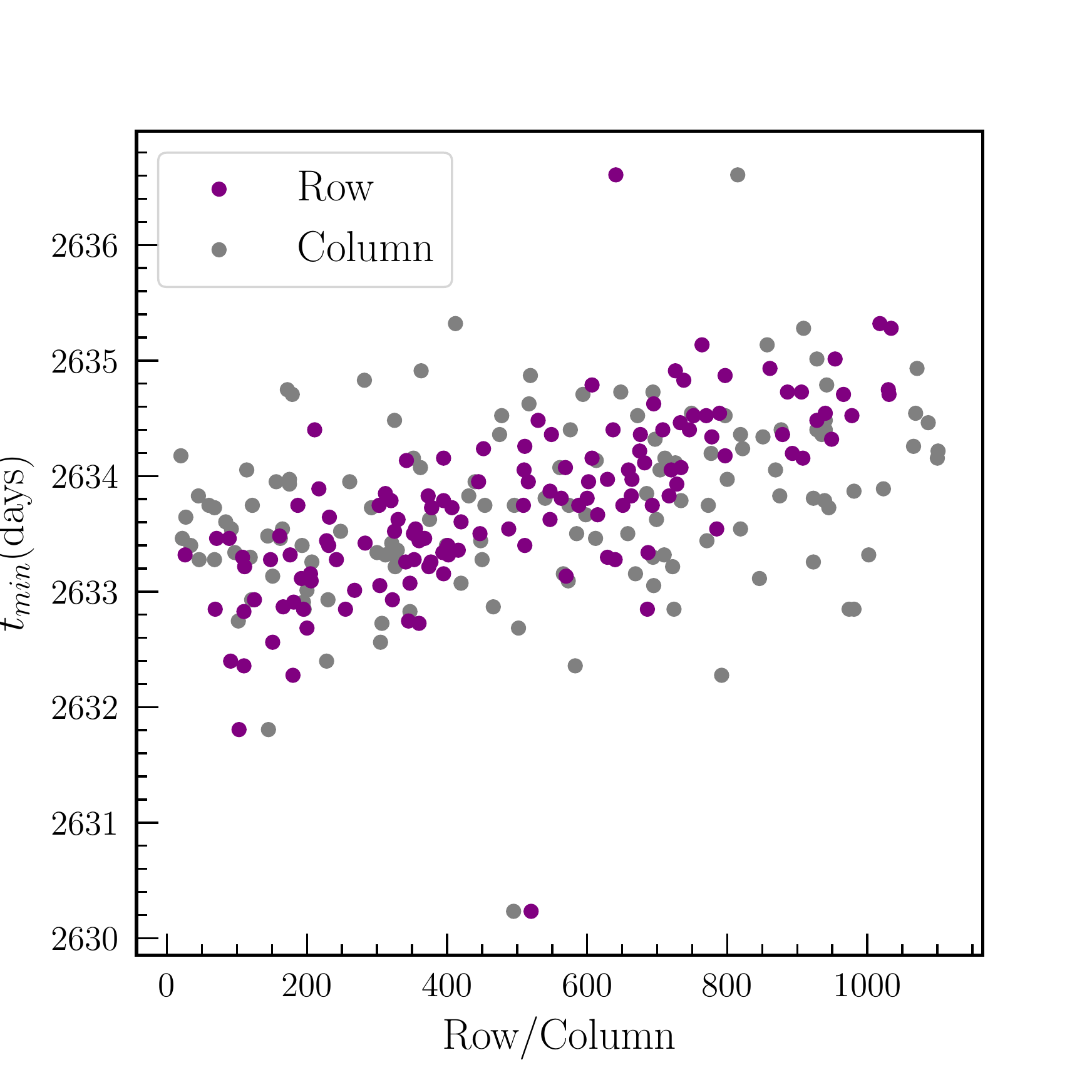}
    \caption{Pixel row and column dependence of a time-shifted dip feature appearing in multiple light curves in channel 33.  {\em Top:} Image of the PSF at the maximum and minimum of the light curve.  {\em Middle:} An example light curve, here from channel 33 (module 11), showing the location of the maximum and minimum features as illustrated in the images.  The vertical lines indicate where we detect the those features and where each of those images were taken.
    {\em Bottom:} Timing of the minimum with pixel row (purple) and column (in grey) for channel 33.   Rolling band is characteristically responsible for a flux increase that propagates across CCD rows. The features indeed shift as a function of pixel row and column, but this trend is a broader function of radius from the center of the FOV as we will illustrate in the following figures.}
    \label{fig:fig6}
\end{figure}

In Figure~\ref{fig:fig6} we illustrate a slightly different pattern affecting another channel (33). We measure the time when a peak and dip {\em pair} occurred in Campaign 8 for all point sources greater than 17th Kepler magnitude. In the top row, the images at the light curve peak and dip rule out neighbors, asteroids, and extreme pointing errors as the origin of these features. In the middle panel, we show that the peak and dip are measured as the maximum and subsequent minimum of the light curve as marked by the blue and orange vertical lines corresponding to the epochs of each image in the top row respectively.  In the bottom panel, we plot the dip epochs, $t_{min}$, as a function of pixel row and column for the full sample of light curves.  The dip feature indeed shifts as a function of pixel row in this particular channel. 
We test for global spatial dependence across the focal plane next by transforming row and column pixel coordinates to radius. The diversity of patterns, for example the number, width and amplitude of peaks and dips, makes it more difficult to measure the time dependence and spatial dependence of such features.

\begin{figure*}
    \centering
    \plotone{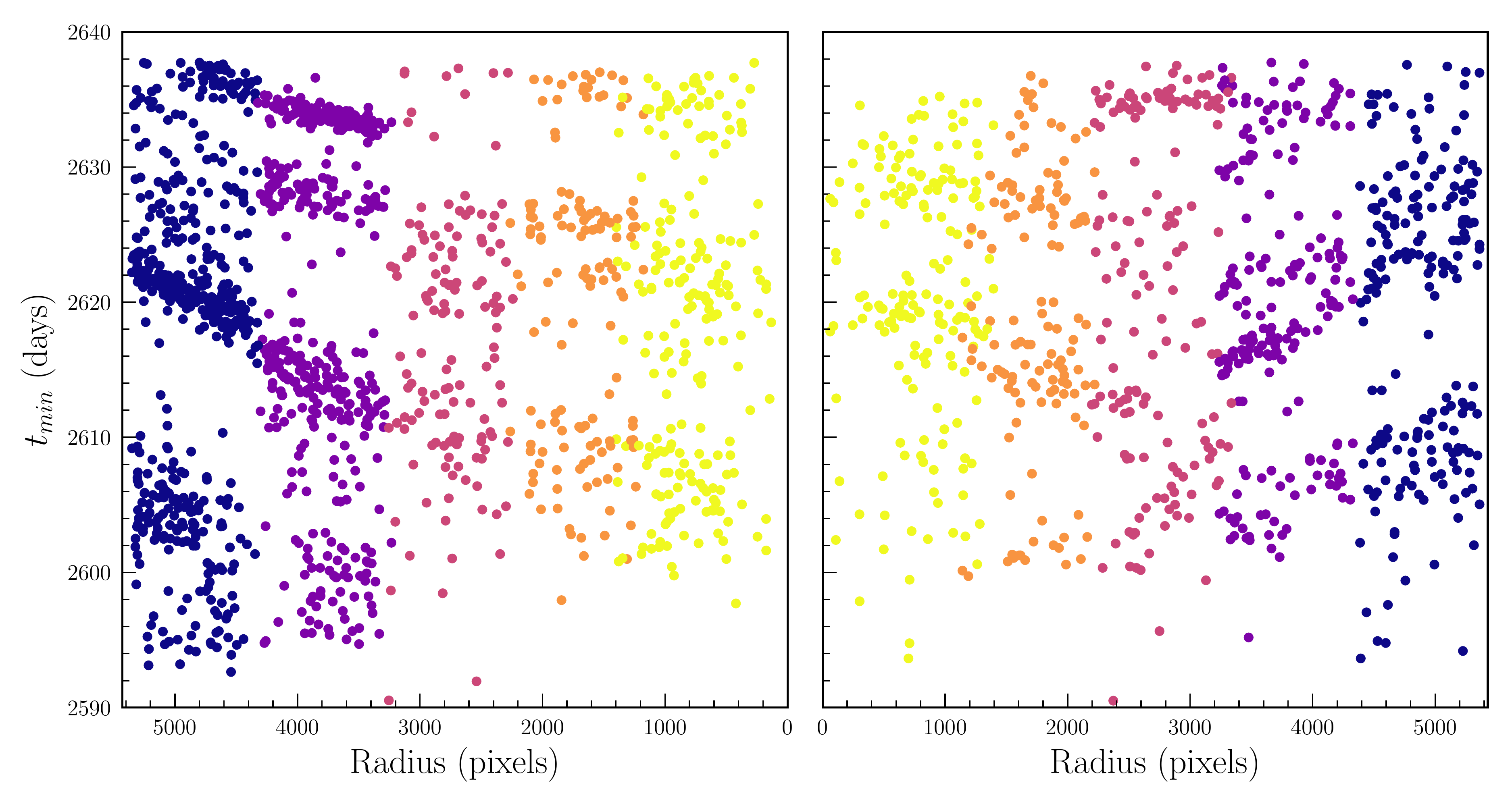}

    \caption{Timing a sequence of two dip features in the light curves of point sources with fainter than 16th Kepler magnitude for all channels in modules 11, 12, 13, and 14.  The scatter plots depict the epochs of two dip features per target as a function of radius from the FOV center.  Channels are colored such that the gradient from light to dark indicates increasing radius. For example, navy points on the left are channels 33 and 36 and on the right are channels 51 and 52.  The lack of randomness indicates that the detected minima/dips are systematic and form trends that cross channel boundaries.  The channels tested correspond to horizontal (x-axis) modules across the FOV diagram in Figure~\ref{fig:fig4}.}
    \label{fig:fig7}
\end{figure*}

\begin{figure*}
    \centering
    \plotone{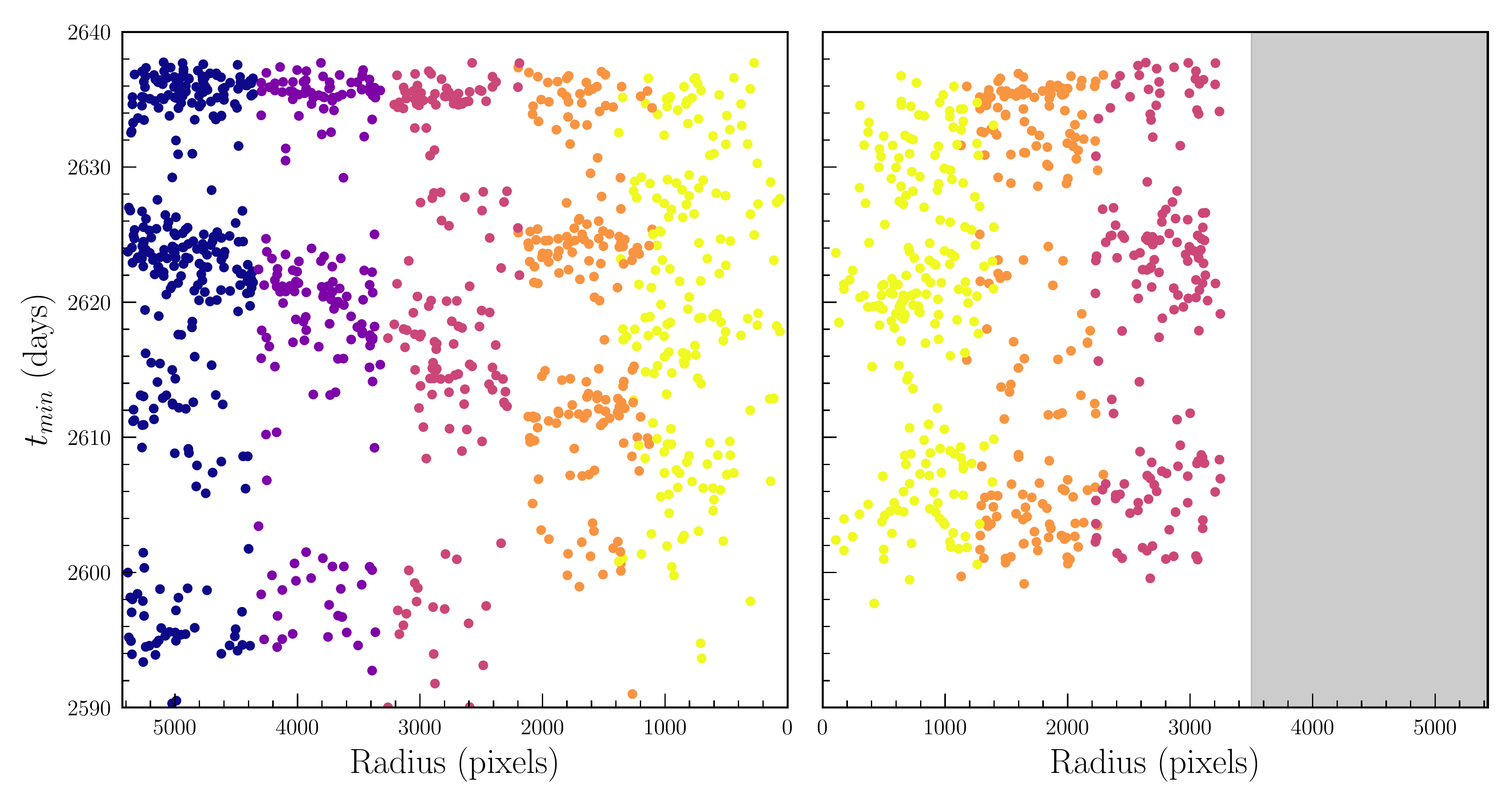}

    \caption{Same analysis as in Figure~\ref{fig:fig7} but for vertical (y-axis) modules 8, 13, 18, and 23 shown in Figure~\ref{fig:fig4}. Failed module 3 (channels 5--8) is shaded in grey. Time dependence and radial dependence of these systematics show the least variance at the outer-most radius.}
    \label{fig:fig8}
\end{figure*}

Figure~\ref{fig:fig4} provides a big-picture view of systematics that vary in pattern and phase across the CCD array with no obvious positional dependence.  Instrumental trends are most likely superposed with astrophysical signal and are harder to characterize from anything but very faint targets, especially if the noise is additive. 
To measure the time-shift of peak and dip sequences such as ``N", ``M", or ``W" shaped features, we use a wavelet algorithm to time the occurrence of several local minima per light curve, similar to our approach in Figure~\ref{fig:fig6} for channel 33. Flux minima or dips due to astrophysical signal are expected to occur randomly in time, whereas any one or more minima that occur at about the same time for several light curves reveal an instrumental origin.

To test if the timing of sequential dips (for example a ``W" feature) exhibit a spatial dependence from channel to channel, we plot the times, $t_{min}$, for the last two dip features per light curve as a function of radius from the FOV center.  Figures~\ref{fig:fig7} and \ref{fig:fig8} show timings of the dip features for light curves observed in modules 11 (which includes channels 33--36 as seen in Figure~\ref{fig:fig4}), 12, 13, 14, and 15 (channels 49--52), which span the full diameter of the focal plane, and perpendicular modules 8 (channels 21--24), 13, 18, and 23 (channels 77--80; short of a full diameter due to failed module 3)\footnote{See https://keplerscience.arc.nasa.gov/the-kepler-space-telescope.html}. 
We account for the CCD rotations and readout directions of the four channels in each module according to rotations noted in Figure 24 of \citet{KeplerHandbook} and we track radius in units of pixel coordinates.  

Figures \ref{fig:fig7} and \ref{fig:fig8} reveal a time dependence and radial dependence for systematics that in fact cross channel boundaries.  Transitioning colors indicate these channel boundaries.
On some channels this approach reveals more than two clusters.  Since clusters point to correlated dips in the light curves of multiple targets, it is possible that the depths of these dips vary across the light curve sample, such that for a subset of light curves the first two dips were the most prominent ``W" shape and for other light curves the last two dips were the most prominent ``W" shape.
Thus, some channels appear to have up to three or four clusters which reflects diversity in the depth width of the instrumental signatures on that channel.  
However uncertain the source of the instrumental signature, it is empirically clear that the signature crosses channel boundaries and cannot be rolling band noise.

\subsection{Magnitude Dependence}
\begin{figure*}
    \centering
    \plottwo{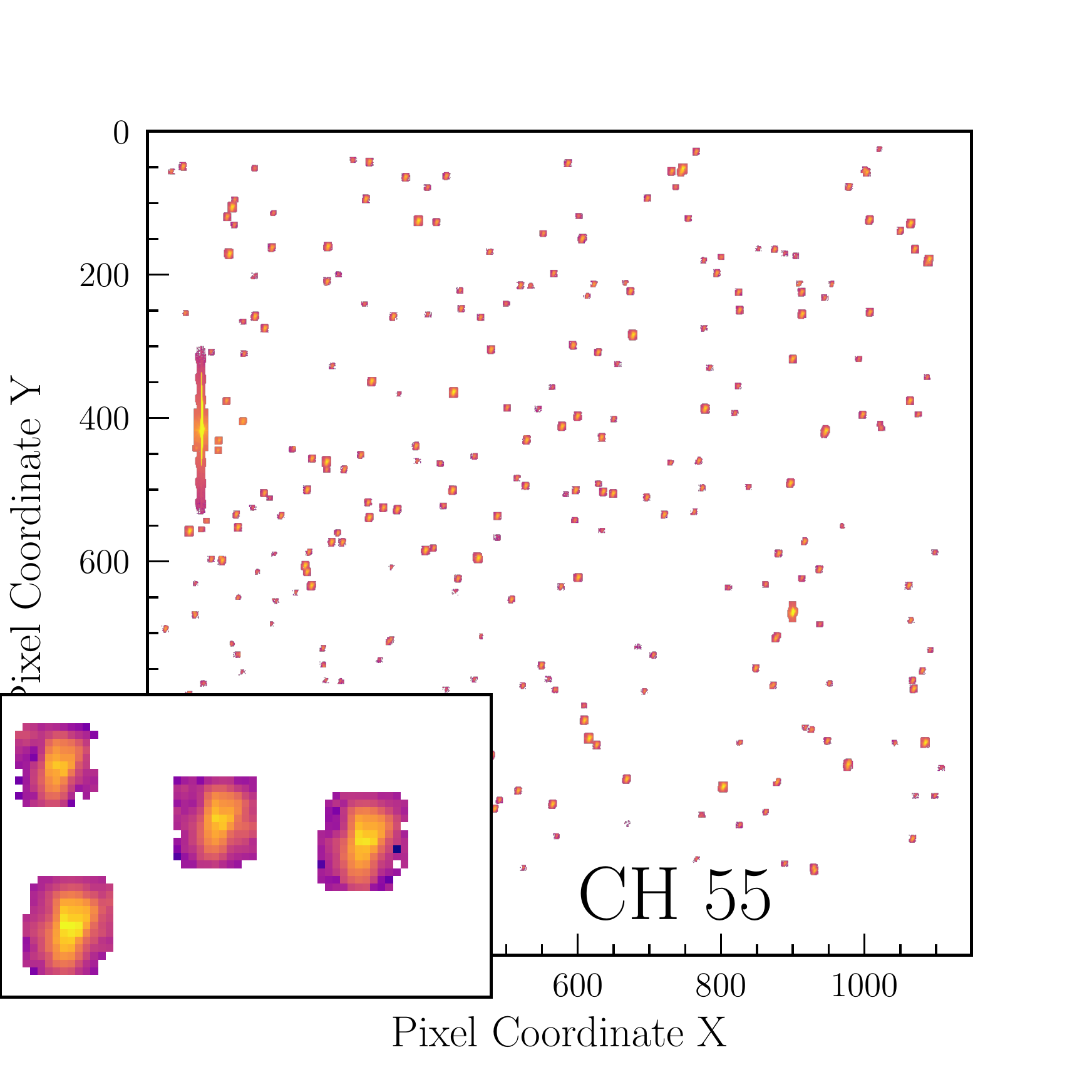}{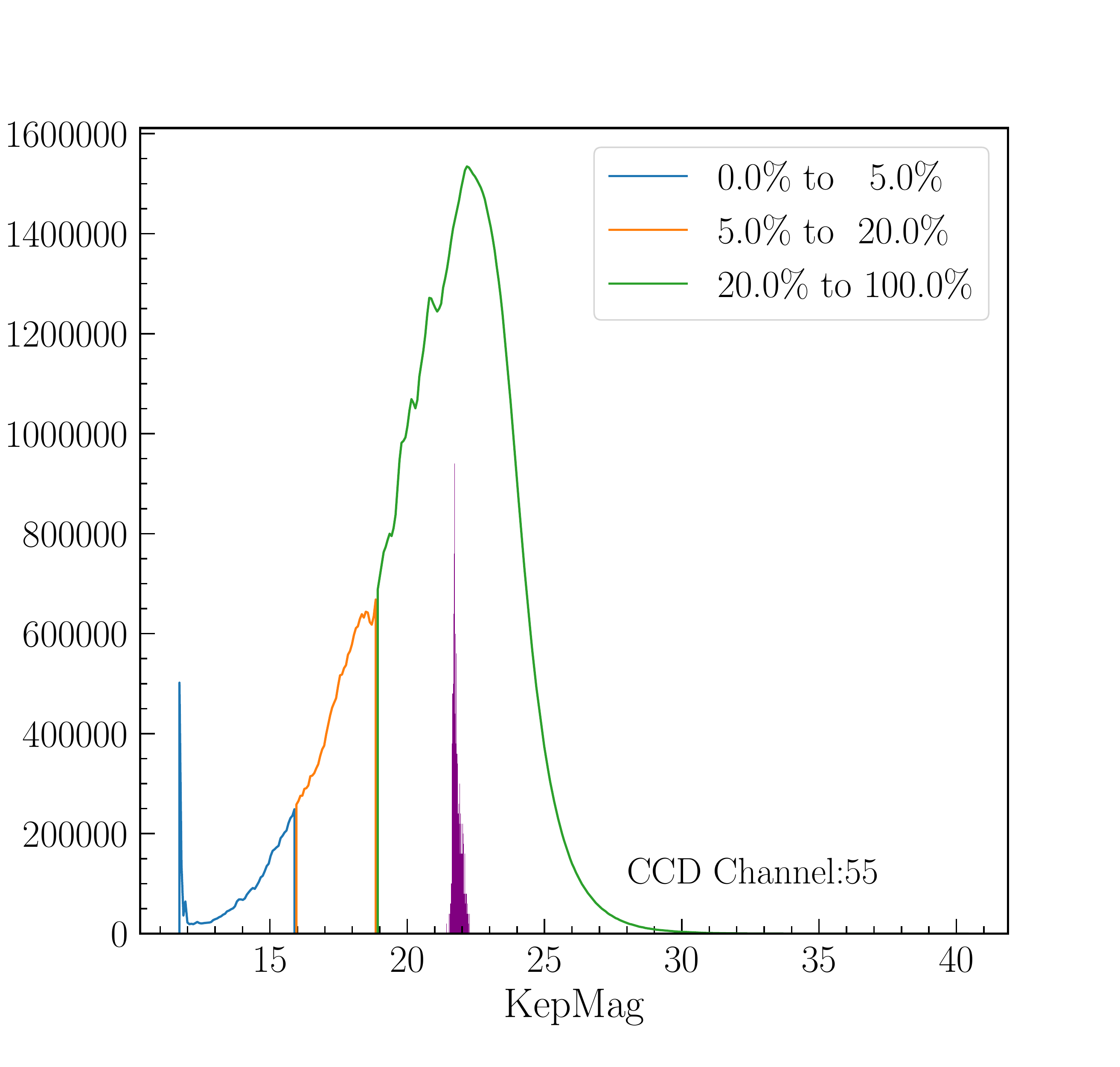}
    \caption{Pixel magnitudes in Campaign 8, channel 55.  In the left panel, we show all epoch-1 postage stamps. An inset that was positioned on a randomly selected group of targets shows that very few pixels are truly empty.  In the right panel, all channel 55 pixels for all epochs are plotted as a magnitude histogram. The distribution of pixels corresponding to the targets shown in the inset is represented by the purple bins (the absolute counts of this distribution are scaled for the benefit of making them visible).  There are very few truly empty pixels that can be used for computing local or global backgrounds.
    }
    \label{fig:fig9}
\end{figure*}

Since it is possible that the systematics are additive and thereby are more dominant in faint sources with low counts, we explored a magnitude dependence at the pixel level and the target level. 
The majority of bright K2 stellar sources reside in the magnitude range $12 < K_{p} <15$, AGNs reside in the magnitude range $15< K_{p}<21$,
and background pixels generally have magnitudes $K_{p} > 18$ depending on the size of the postage stamp and the extent of the PSF.  Faint background sources $\sim 19$th magnitude may also reside in every pixel \citep{KeplerHandbook}.

In Figure~\ref{fig:fig9}, we selected a worst-case example, channel 55 (see Figure~\ref{fig:fig4}), from Campaign 8 to investigate the typical distribution of pixel illumination, measured in magnitudes.  We show the distribution of Kepler magnitudes for individual pixels in order to quantify the percentage of ``empty" pixels that may be used to characterize instrumental trends in each CCD background.  The left panel includes an inset with example postage stamps at a snapshot in time, demonstrating that the postage stamps contain very few truly empty pixels that can be used for computing local backgrounds.  The raw magnitude distribution of pixels in the inset is also shown in the right panel by the purple bins.  Pixels in these four postage stamps range from $21-22.2$ Kepler magntiudes.  Although nearly $40-50\%$ of pixels are fainter than the faintest pixels in these four postage stamps, these faint sources likely have no background pixels and their PSFs may even jitter out of the cutout boundaries.

\begin{figure*}
    \centering
    \epsscale{1.15}
    \plotone{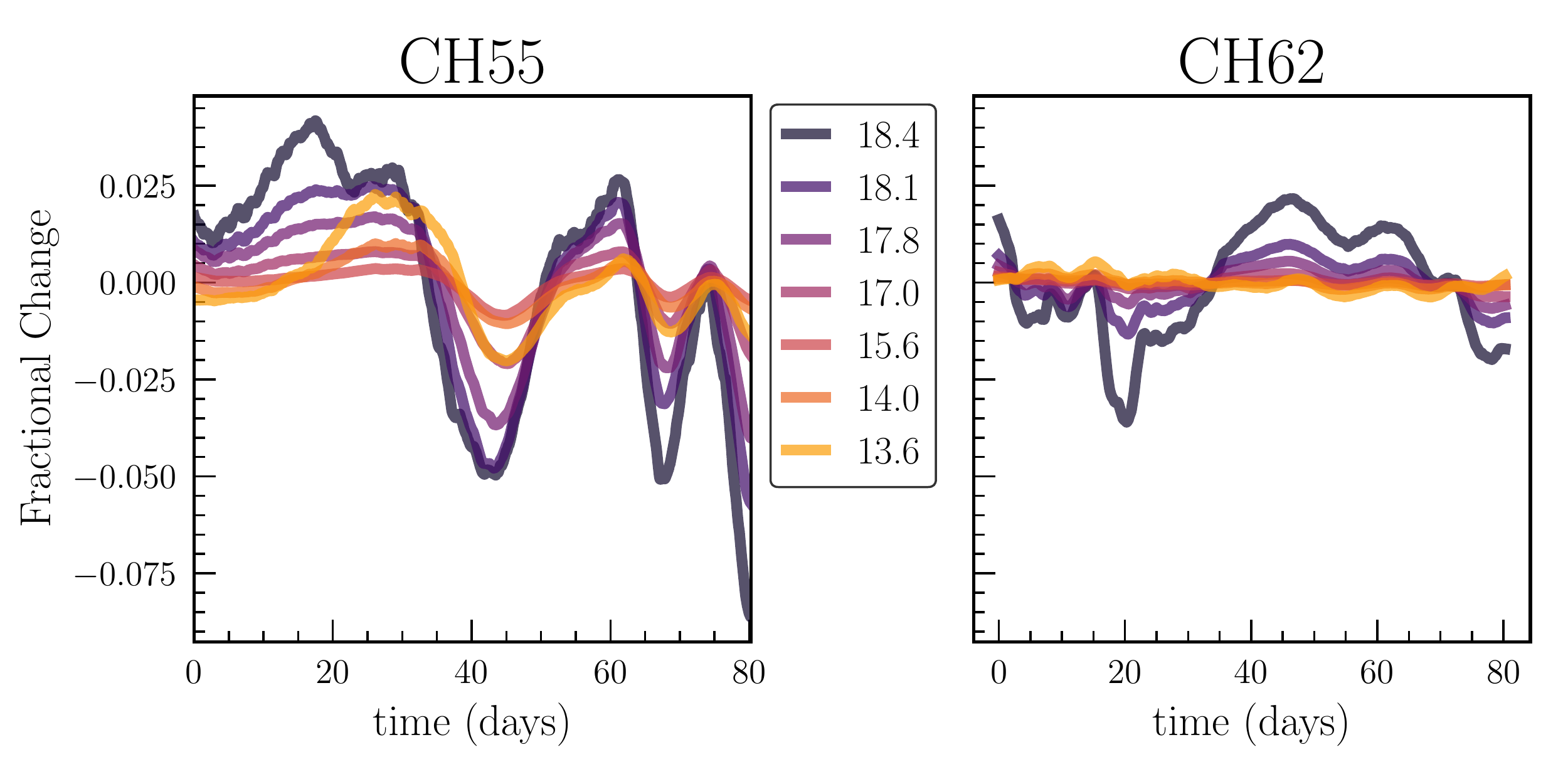}
    \caption{Target magnitude dependence on channel 55 (edge of FPA) and channel 62 (center of FPA) in Campaign 8. Each panel shows flux-percentile target light curves in units of fractional change.  Percentile light curves are labelled by their corresponding mean target magnitude.  A 13th magnitude target on channel $62$ has noise fluctuations with amplitudes less than $1\%$. On the other hand, in channel 55, targets fainter than magnitude 18 regularly have multiple dips at the fractional amplitude of $\sim 5\%$ or greater}.  
    \label{fig:fig10}
\end{figure*}

To examine how systematic effects reveal themselves differently as a function of target magnitude,
we compute and stack full-postage-stamp flux light curves and select the 5th, 10th, 15th, 25th, 50th, 75th, and 85th percentile at each epoch.  In Figure~\ref{fig:fig10}, we plot these percentile light curves in units of fractional change and label each with its corresponding mean (target) magnitude for channel 55 (left panel; edge of focal plane) and channel 62 (right panel; center of focal plane).  Figure~\ref{fig:fig10} reveals that fainter magnitude targets have larger fractional changes than brighter targets, especially on bad channels.  Channel $55$ shows that targets with Kepler magnitudes fainter than $18th$ exhibit amplitudes greater than $5\%$ while on better channels, such as $62$ located at the center of the focal plane array (FPA), faint targets exhibit fractional changes at $\sim 2\%$ or better.

\begin{figure*}
    \plottwo{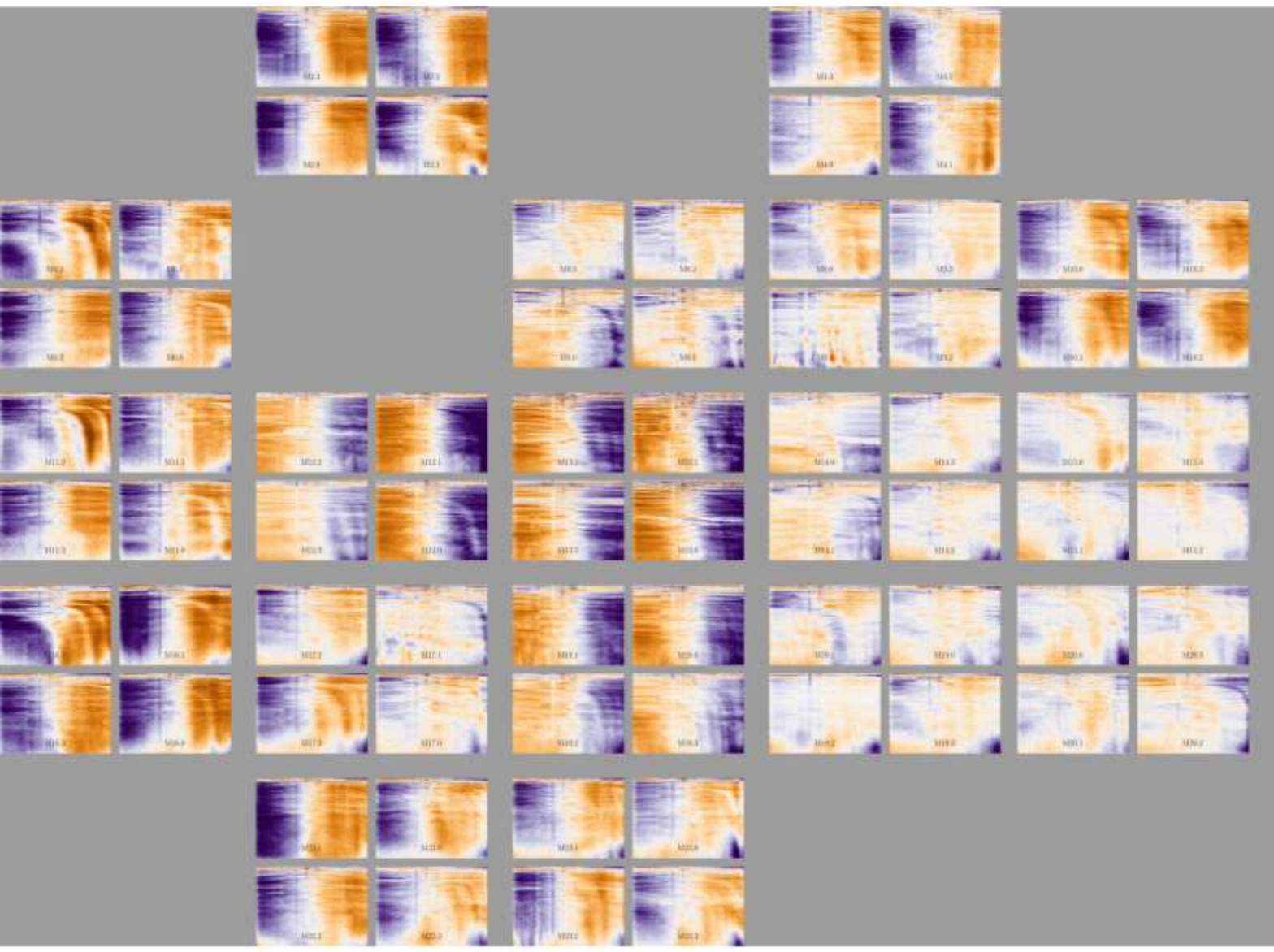}{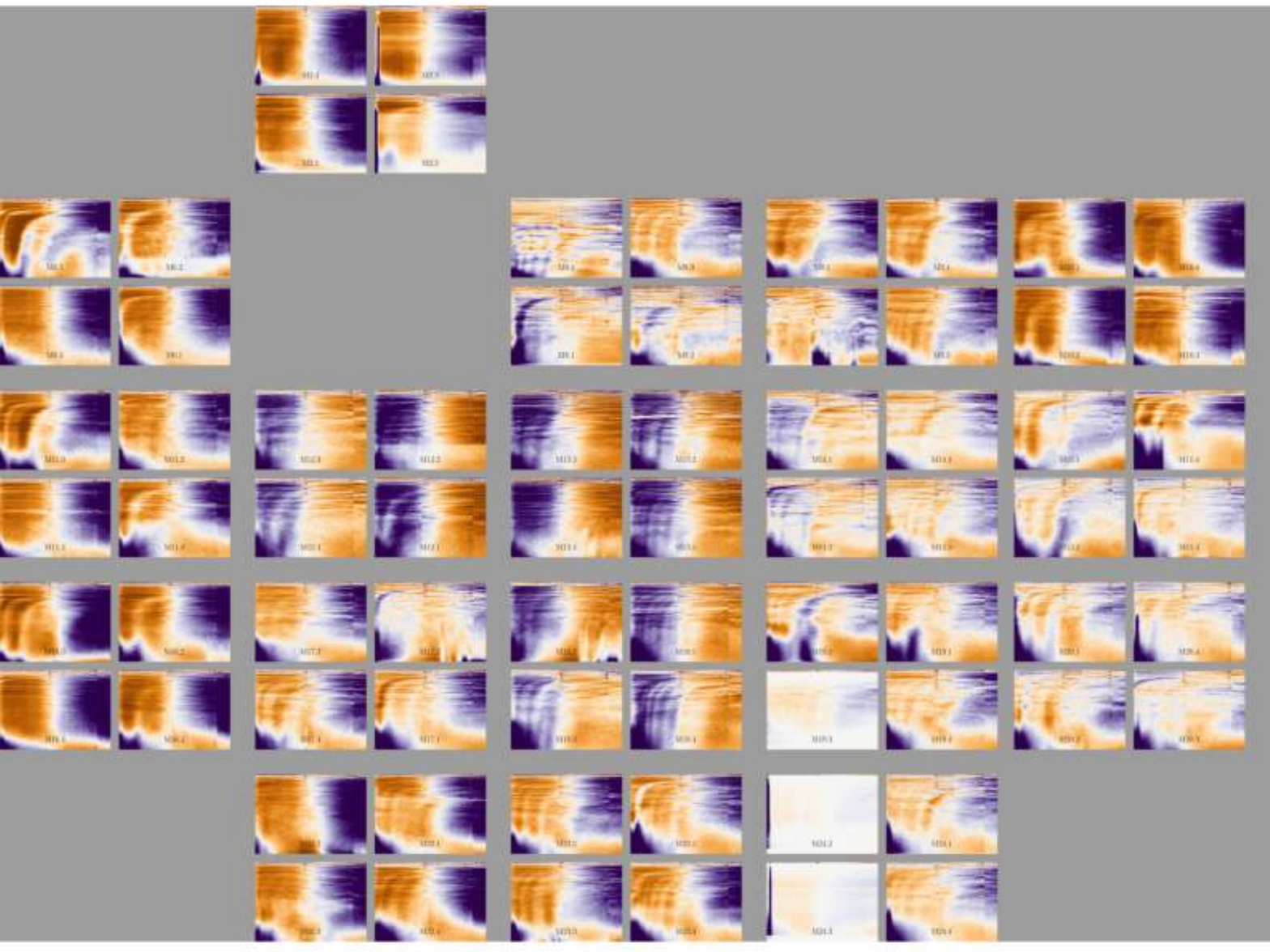}
    \caption{
    Pixel-correlation-intensity maps as a function of magnitude. Left: Campaign 8; right: Campaign 16.  Magnitude-dependent systematics in C8 are seen to be reversed in time for C16. The $y$-axis of each channel corresponds to magnitude bins. Each row, or magnitude-binned background light curve is generated from an averaging of background pixels.  Each row on the intensity map is a magnitude-specific global background, stacked together to visualize systematics across the CCD array as a function of magnitude.  Purple corresponds to normalized flux above the mean. Orange to purple indicates an increasing trend in each magnitude-binned light curve. 
    Fainter magnitudes (towards the bottom of each channel) exhibit more complicated systematics. }
    \label{fig:fig11}
\end{figure*}

We display the results of a related analysis for {\em all} of the channels (from Campaigns 8 and 16) in Figure~\ref{fig:fig11}, where
correlated systematics can be seen across the K2 FOV.
In this presentation we have stacked 500 (magnitude-binned) median light curves generated from ``empty" pixels (the faintest 80\% shown in Figure~\ref{fig:fig9}) to produce intensity maps for the entire CCD array for Campaigns 8 and 16 using the method developed by \citet{jack2018}. Non-random structure in these intensity maps show pixel correlations (in time; x-axis in each channel) due to the instrument as a function of magnitude (y-axis in each channel).  
We applied flux smoothing and whitening (transformed to having a mean of zero and a variance equal to one) in order to uniformly compare the strength of systematics across magnitudes and channels.  We use a divergent purple-to-orange colormap to represent trends above and below the median flux value, respectively. Saturated colors from purple to orange indicate a decreasing trend in each magnitude-binned light curve as seen in Campaign 8. For example, shifting dips show up as the orange Florida shape (channel 15) or striping feature (channel 55) in the left-most upper block in Figure~\ref{fig:fig11}.  Trends in Campaign 16 are seen to be reversed in time as compared to Campaign 8 for corresponding channels.

In all campaigns (not shown here), modules 14, 15, 19, 20 are least affected by systematics (i.e., intensity maps contain the most white), whereas modules 6, 11, 16 (on the left edge of Figure~\ref{fig:fig11}) exhibit the strongest intensity systematics. Scanning each intensity map from the top-down, the direction of increasing magnitude (dimmer pixel illumination), we observe structure corresponding to peaks (purple) and dips (orange) in median light curves. Strikingly, all of this correlation or pattern structure comes from pixels with magnitudes from $19-30$ which are used for local background subtraction.   
We attempted to use these magnitude-binned background light curves to correct various stellar and AGN light curves, but it was unclear how to evaluate an optimal fit without suffering the same failings of a PCA-approach. The magnitude and spatial systematic dependencies discovered in these analyses must be considered during aperture section and arc-drift removal steps.

Three inner modules exhibit a trend in Figure~\ref{fig:fig11} that is opposite to the rest of the modules in the array.  Modules 12, 13, and 18 exhibit a negative slope, purple to orange in Campaign 8, while all other modules have a positive trend, orange to purple.  This result provides evidence that focus is a dominant source of error responsible for long-term tends. Given the spacecraft orientation in Campaign 16, for example, outer modules may have started the campaign in better focus and ended out of focus with a broader PSF, while inner modules may have had the opposite trend with a narrowing PSF or vice-versa.

\begin{figure*}
    \plotone{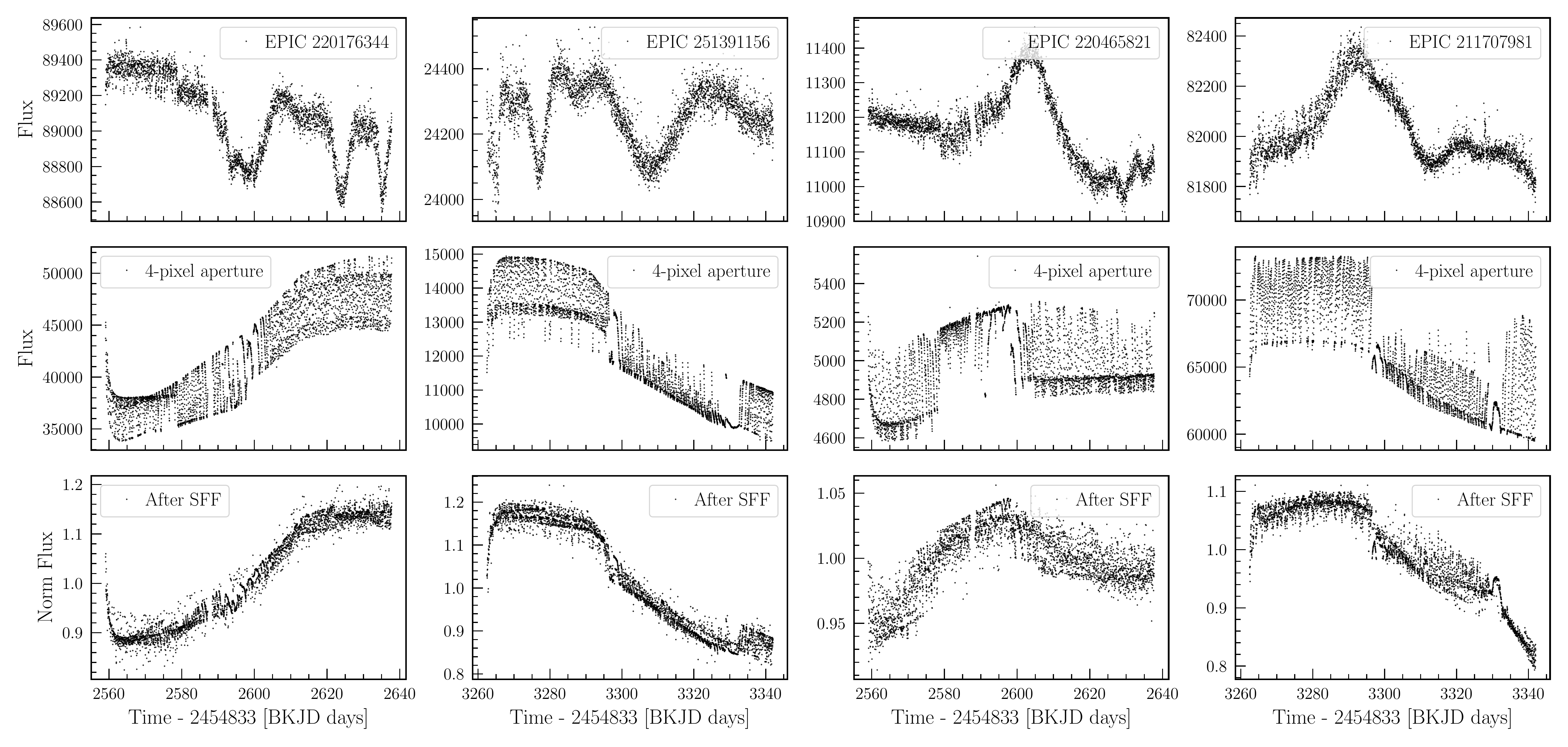}
    \caption{Custom SFF reprocessing examples. Each column is a unique target from Campaign 8 (1st and 3rd columns) or 16 (2nd and 4th columns). 
    Each row is a step in a typical extraction procedure. 
			\emph{top row:} Raw light curves taken from the complete postage stamp. \emph{middle row:} Raw light curves taken from the 4 brightest pixels in the postage stamp at any given time. 
			\emph{bottom row:} ({\tt lightkurve}) SFF light curve using 4 brightest pixels as aperture.
			The decrease in aperture size significantly increases arc drift which is ``corrected" by the application of SFF. The smaller aperture removes the long-term systematics we see, but the light curve is still dominated by the effects of arc drift, even after SFF (as is clearly seen by the ``mirrored" nature of the light curves illustrated here and in Figure~\ref{fig:fig4}).}
    \label{fig:fig12}
\end{figure*}

\subsection{Optimizing the Aperture}

Aperture photometry is a technique that assumes a useful shape such as a square or circle rather than a PSF model to collect and sum observed counts.  After estimating the total counts from all pixels within some defined area, an estimate of the background can be subtracted to compute the flux of the source.  An optimal aperture captures as much of the signal (as close to 100\% of the true PSF) as possible to maximize S/N.  Fainter sources typically require larger apertures.  However, as the number of pixels increases, the PSF wings of a faint source have increasingly larger errors. These considerations are particularly relevant for AGN in crowded fields and local (low luminosity) AGN that do not significantly outshine their host galaxy. In such cases, flux extraction of the faint AGN becomes prohibitively more difficult since the outer wings of the PSF are convolved with the extended galaxy profile.  The K2 systematics discussed in the previous subsection present a tertiary motivation for PSF modelling or developing a careful aperture optimization scheme.

Careful frame-by-frame aperture photometry may be a way to mitigate time-dependent focus changes in K2 light curves.
Empirically, we find that the aperture size highly influences the strength of the instrumental signature under investigation.
Typically for AGN, we would want very small apertures to reduce blended background sources in dense fields. We demonstrate the issue with this approach in Figure~\ref{fig:fig12} using the {\tt lightkurve} package \citep{lightkurve2018}\footnote{{\tt lightkurve} also contains functions for centroid position information and PSF engineering data.}.

\begin{figure*}
    \plotone{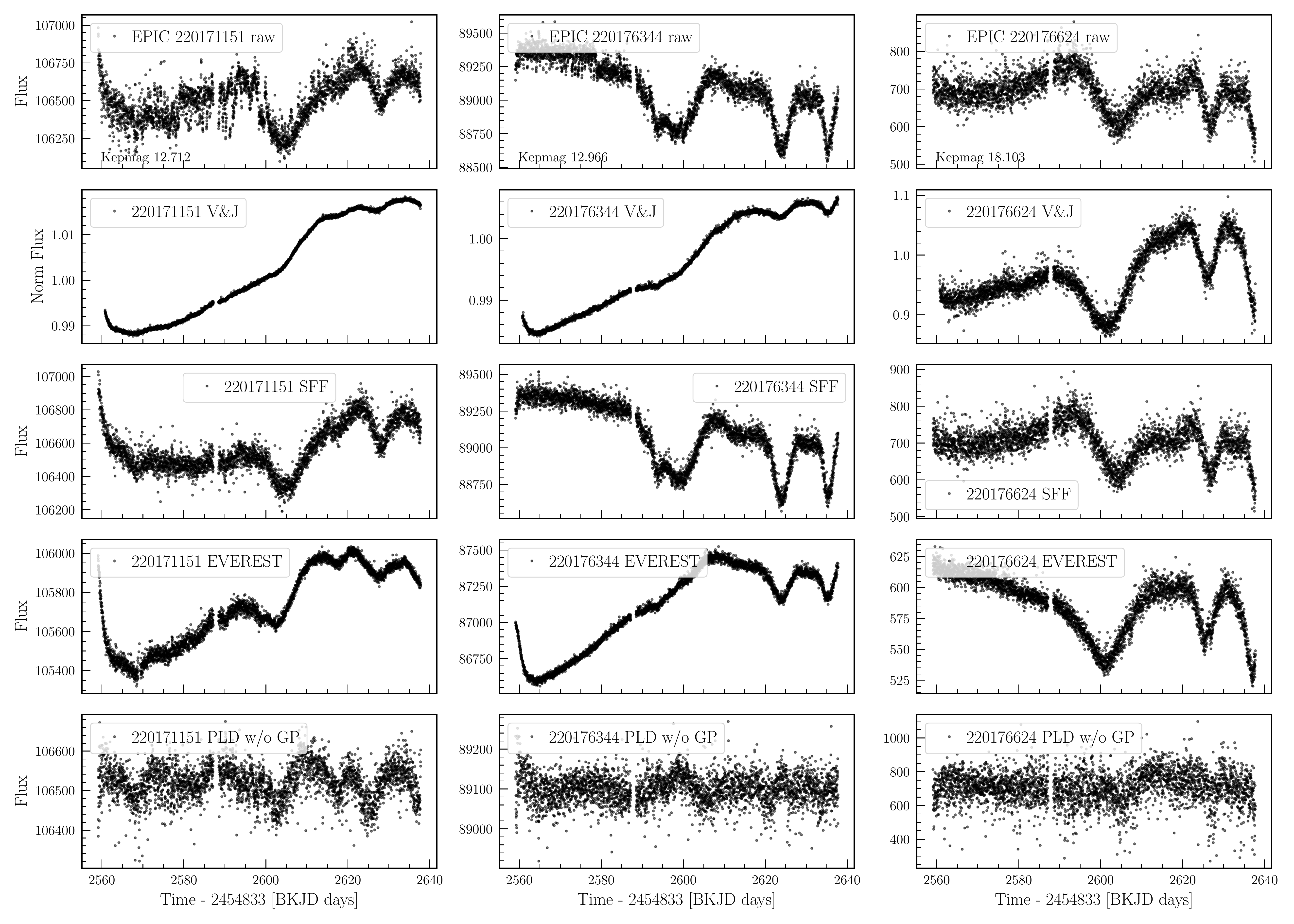}
    \caption{PLD and SFF corrections from the python {\tt lightkurve} package (without background subtraction) applied to three objects from campaign 8, channel 55. Each column is a unique target.
		\emph{top row:} Raw light curves extracted from the complete postage stamp.
		\emph{second row:} VJ14 SFF light curve using the full postage stamp. 
		\emph{third row:} ({\tt lightkurve}) SFF light curve using the full postage stamp. \emph{fourth row:} ({\tt EVEREST} 2.0) PLD + GP process corrected light curve using the full postage stamp.
		\emph{bottom row:} ({\tt lightkurve}) PLD extracted light curve without GP kernel.
		Instrumental systematic effects persist after arc-drift removal in all rows.}
    \label{fig:fig13}
\end{figure*}

We observe a trade-off between arc-drift correction performance and the inclusion of complex systematics that are difficult to detrend. In Figure~\ref{fig:fig12}, we compare 4-pixel moving aperture light curves for four unique targets (columns) to full postage-stamp light curves (little arc drift but maximal error contributed by more pixels).  In the first row of Figure~\ref{fig:fig12}, we plot full postage-stamp light curves for pairs of objects with mirrored systematics in Campaigns 8 and 16.  In the second row, we show the 4-pixel moving aperture for each object. Finally, in the third row, we show the SFF-corrected moving aperture light curves.  The smaller aperture minimizes the inclusion of instrumental trend, as seen in the second row, but the light curve is still dominated by the effects of arc drift, even after the application of the SFF correction in the third row.

Arc-drift requires larger apertures to be properly corrected, but because Kepler pixels are large this ultimately includes more sources of error. An additional difficulty for aperture photometry and PRF modelling using commissioning data \citep{Bryson2010} is that the true PSF may change with time due to temperature dependent focus changes throughout the course of a campaign as we have been suggesting.  A fixed aperture size, although moving to mitigate arc-drift, would fail to preserve a constant percentage of the true PSF as it breathes in response to temperature changes experienced as a consequence of the Sun-spacecraft orientation.

PSF modelling may be a crucial part of the error mitigation process for faint targets at the largest distance from the center of the FOV.  However, is unclear how dramatic PSF changes may be throughout a campaign and if they are radially dependent given initial focus non-uniformities.  It is also possible that the PSF may go out of focus to the point that the signal of interest spreads beyond the edges of a postage stamp.  There is no easy way, to our knowledge, to deal with these cases.

\section{K2 Re-processing Pipelines}
\label{sec:software}
Two of the most successful and popular arc-drift-corrected K2 data sets available through \textit{MAST} are the self-flat fielding (SFF) method of \cite{VJ2014} (henceforth VJ14) and the pixel level decorrelation (PLD) method implemented in {\tt EVEREST} by \citet{Lugar2018}. Both of these methods have also been implemented in the {\tt lightkurve} package \citep{lightkurve2018}. In this section, we present and discuss SSF and {\tt EVEREST} 2.0 public light curves, showing that these data products are successfully free of arc-drift, but still require instrumental noise removal at low to mid-range frequencies prior to employing time series analyses such as estimating power spectra.

The SFF method corrects for systematics correlated with centroid arclength (drift) by fitting a third order B-spline to position offsets over windows of $\sim1.5$ days across the light curve.  VJ14 light curves are plotted in the second row of Figure~\ref{fig:fig13} (for multiple Kepler magnitudes). For comparison, raw light curves are shown in the first row.  The VJ14 pipeline first defines an optimal stationary aperture by defining a circle of pixels around the target centroid where radius is a function of magnitude. A median background is estimated from excluded pixels and subtracted from the optimal aperture light curve.  Multiple circular and pixel response function based apertures are available in \textit{MAST}. \textit{K2 SFF} light curves are continuum normalized and no flux errors are available.  A flexible implementation of the SSF method is available in the {\tt lightkurve} package.

Pixel level decorrelation (PLD; \citealt{Deming2015}) is another highly successful method for arc-drift removal \citep{Lugar2018}. PLD is a Taylor-series expansion of fractional flux changes in individual pixels and pixel products intended to model covariances  in flux variations (across pixels) resulting from instrumental motion. In {\tt EVEREST} 1.0, PLD is calculated from pixels within each individual target postage stamp. In {\tt EVEREST} 2.0, PLD weights are computed from an ensemble of neighboring stars on the same channel (limited to bright stars in the magnitude range of $11< K_{p}<13$) and thus is referred to as the neighboring-PLD (nPLD).  {\tt EVEREST} 2.0 also makes use of a Matern 3/2 Gaussian process (GP) kernel to account for systematic covariance between cadences when fitting the noise model. 
The PSD of a Matern 3/2 kernel has an exact closed-form representation, with power law slope $-4$ in the limit of high frequencies and constant for low frequencies, with a knee set by the natural frequency tuning parameter $\omega_0$ \citep{Celerite}.  The choice of a Matern 3/2 in {\tt EVEREST} may therefore imprint an artificial power law slope into the observed PSD.  We experimentally verified that these power law slopes appear in finite sample sizes by drawing random sequences from Gaussian Process kernels with known input $\omega_0$ and time sampling comparable to the K2 lightcurves\footnote{https://github.com/BrownDwarf/probabilisticAGN}.
{\tt EVEREST} 2.0 light curves are plotted in the fourth row of Figure~\ref{fig:fig13}.

\begin{figure*}
    \plotone{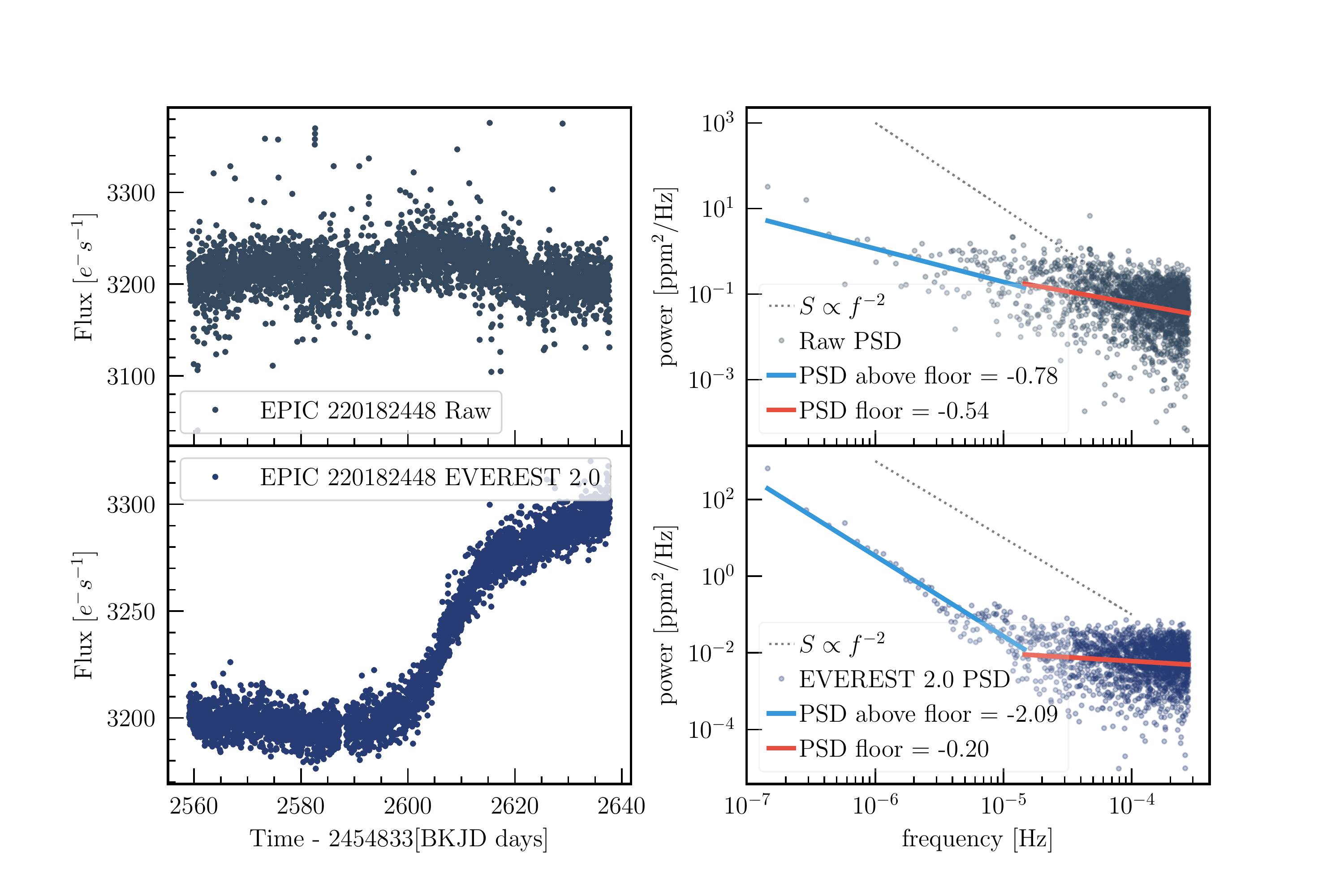}
    \caption{{\tt EVEREST} 2.0 applied to a quiet star. The top left panel is the raw light curve, using all of the pixels on the postage stamp as the aperture. The bottom left panel is the light curve resulting from PLD method and a Matern $3/2$ Gaussian process applied using the same aperture. The right panels are the corresponding PSDs. The black dotted line represents the PSD slope of a damped random walk (DRW or red noise).  The Matern $3/2$ kernel adds in variability that results in a PSD slope that artificially mimics a DRW.}
    \label{fig:fig14}
\end{figure*}

\begin{figure*}
   \plotone{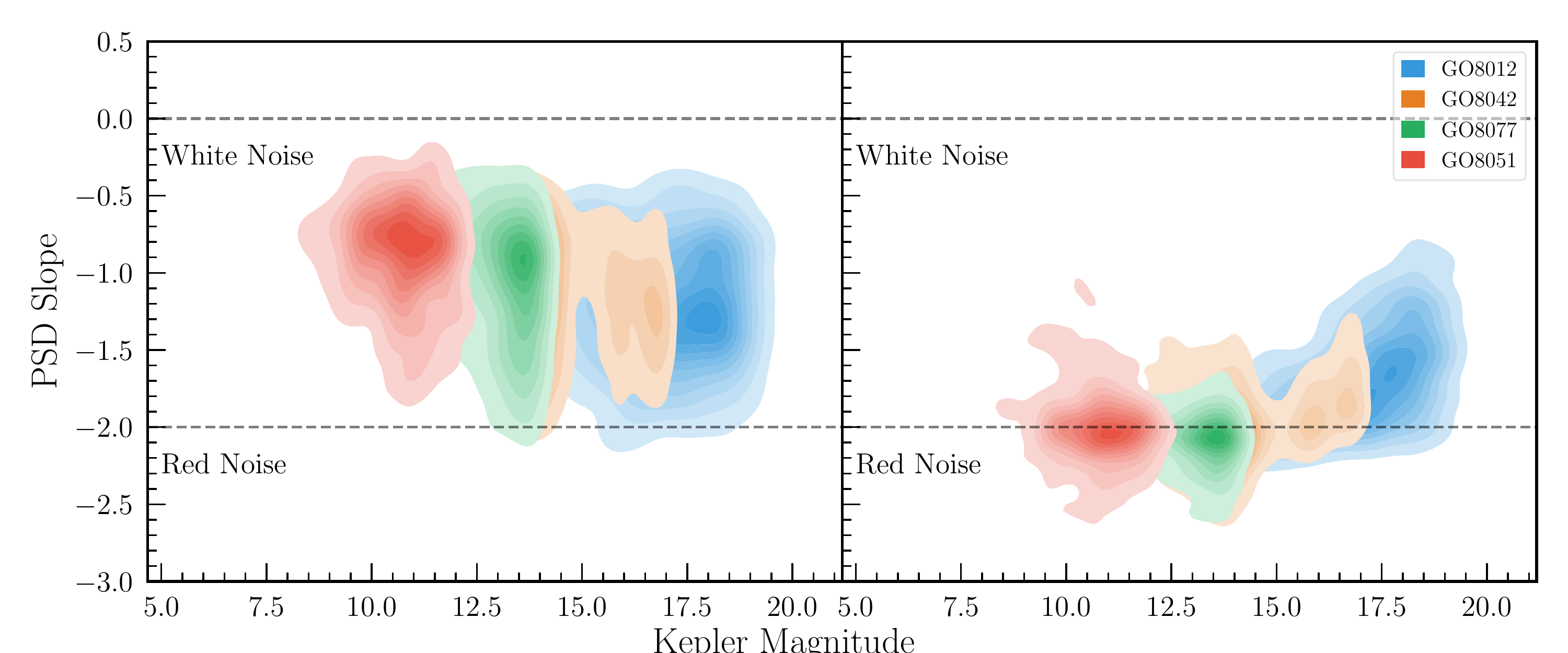}
    \caption{PSD logarithmic slopes as a function of magnitude for non-AGN objects in Campaign 8 before (left panel; raw full-postage stamp light curve PSDs) and after (right panel) {\tt EVEREST} 2.0 reprocessing using the Matern $3/2$ kernel. Here we see that a naive use of the Matern 3/2 kernel artificially steepens the PSD slopes of $9483$ non-AGN targets (from various Guest Observer proposals) to the value of $-2$ which corresponds to red noise spectra.  An incorrect GP model selected for AGN will similarly produce an artificial PSD slope.}
    \label{fig:fig15}
\end{figure*}

Naive use of the Matern 3/2 kernel results in a light curve with added variability at low frequencies (red noise).
In Figure~\ref{fig:fig14}, we show an example raw light curve with little variability for which use of the Matern 3/2 kernel results in added power at low frequencies. In the left panels, we see the raw light curve with arc drift (top) and the {\tt EVEREST} 2.0 light curve (bottom) available through \textit{MAST}. In the right panels, we plot the power spectral density (PSD) estimated with the Lomb-Scargle periodogram and a power law model fit above and below the noise floor.  The PSD slope of the raw light curve resembles white noise with a slope approximating zero, while the PSD of the {\tt EVEREST} 2.0 reprocessed light curve exhibits a slope equal to that expected for AGN variability (PSD $\propto f^{-2}$), but this variability is not real. 

In the case of AGNs and low frequency stellar variables such as M dwarfs, artificial power is added by the use of the Matern 3/2 kernel.    
In Figure~\ref{fig:fig15}, we show the slope of the power spectrum  in {\tt EVEREST} 2.0 lightcurves as a function of {\em Kepler} magnitude.  
Nearly all of the targets, represented by density contours, in Figure~\ref{fig:fig15}, need to be reprocessed with the {\tt EVEREST} 2.0 package or {\tt lightkurve} PLD routine with the Matern 3/2 amplitude parameter set to zero (or GP modeling toggled off) in order to benefit from the n-PLD method.
See also \citet{Saunders2019} who performed an assessment of  the impact of different detrending algorithms on the PSD and determined that PLD with a careful choice of kernel is a promising method for recovering signals that can be modeled by a DRW.

For AGN short timescale variability (and K2 data in particular), we are most interested in estimating the high frequency slope of the empirical PSD. GP processes require prior knowledge of an appropriate model (kernel) for the astrophysical signal. However, {\em Kepler}/K2 observes a unique timescale regime that has never been studied or surveyed ever before for a complete AGN sample. Additionally, the {\em Kepler}/K2 30-minute cadence makes the empirical PSD the best tool for informing our stochastic model selection for long-timescale ($>100$ day) variability studies.  Thus there is a strong desire to rehabilitate the light curves of AGNs observed during the K2 campaigns as discussed in the next section.

\section{Why Rehabilitate K2 Data?}
\label{sec:why}

\begin{figure*}
    \plotone{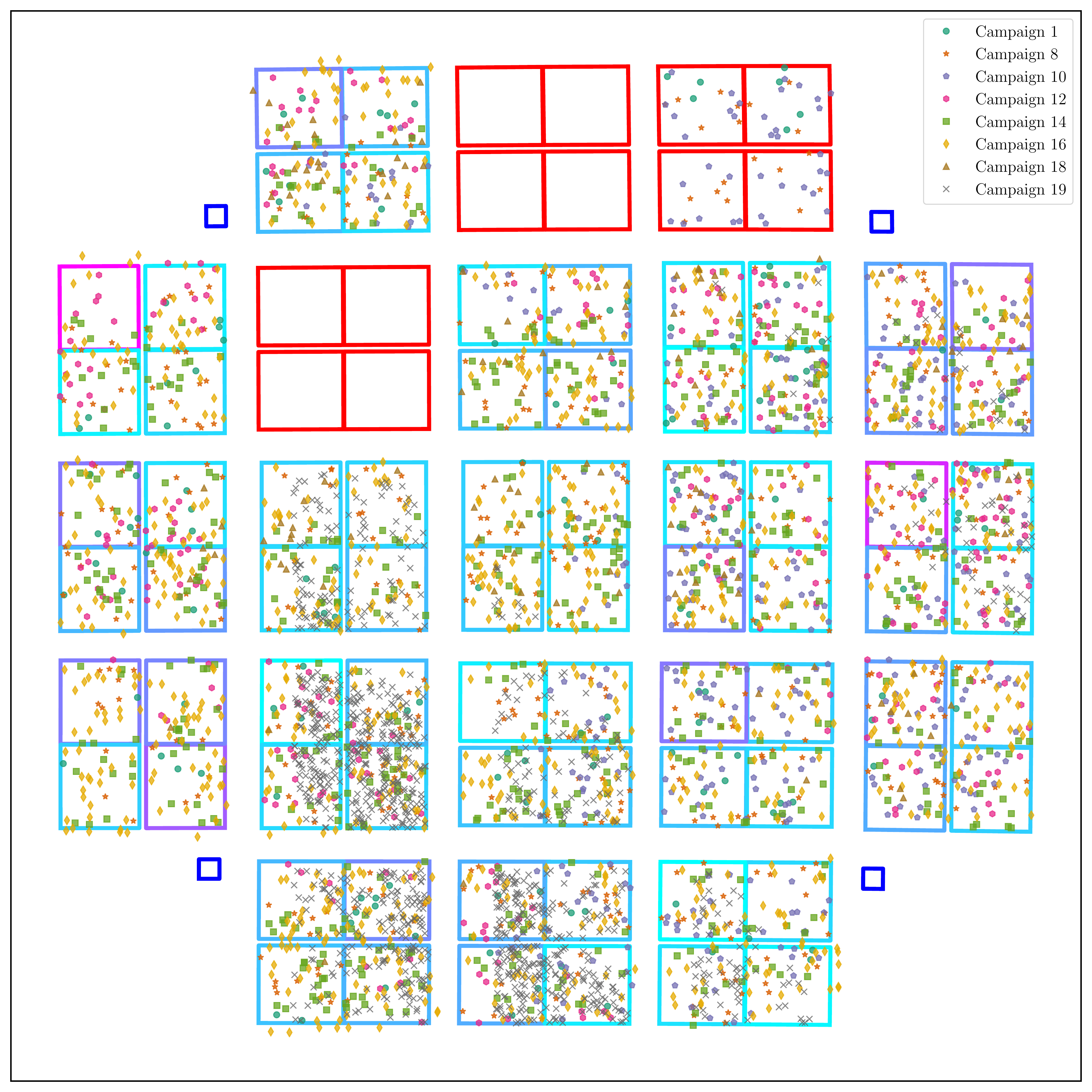}
    \caption{The approximate locations of AGN relative to the K2 field of view. The translations of different campaign fields of views to overlap added in slight rotation error in the AGN positions.  The coloring of the channels is based on the median absolute deviation (MAD) of the median channel light curves (after subtracting a linear trend). The MAD values are calculated from the smoothed median channel light curves shown in Figure \ref{fig:fig4}. The red outline indicates a bad module. Module 4 failed after Campaign 10. The cluster of objects in Campaign 19 are in SDSS ``stripe 82" \citep{York2000}.
    }
    \label{fig:fig16}
\end{figure*}

{\em Kepler}/K2 AGNs may offer significant insights to the overall understanding of AGN variability acquired from investigations of ground-based light curves. PSD (and structure function; SF) features at timescales less than 50 days have been debated in multiple investigations of ground-based light curves.  Short-timescale AGN variability is difficult to disentangle from the effects of gappy, irregular cadences using methods such as the Lomb-Scargle periodogram and structure functions. K2 will be able to reveal if the AGN PSD slope in the high frequency regime differs significantly from the low frequency regime constrained with ground-based surveys.

The simplest stochastic model for simulating optical AGN variability is the damped random walk (DRW). The DRW is, however, an inflexible model with a fixed PSD power law shape (PSD $\propto \nu^{-2}$). 
Empirical PSD and SF analyses of {\em Kepler} AGN from the mission's first phase showed a range of PSD shapes (curvature and slope transitions) (\citealt{Mushotzky2011}; \citealt{Vish2015}, \citealt{Vish2017}; \citealt{Smith2018}). Rehabilitating K2 light curves would provide the opportunity to characterize the high-freqency PSD slope for a large statistical sample of AGNs. In Figure~\ref{fig:fig16}, we show 4000 AGN observed in Campaigns 8, 10, 12, 14, 16, 18, and 19 over-plotted on the Kepler CCD array.

The (relatively) high frequency regime of AGN-PSDs can be investigated as a function of luminosity, black hole mass, and accretion rate. The characteristic frequency or timescale for a slope transition may also be related to physical scales or black hole mass as is the case for AGN X-ray variability \citep{UttleyPSD}. Recently \citet{Simm2016}, \citet{Caplar2017}, \citet{Zinn2017}, \citet{Smith2018} and Moreno et al (2021, in prep.) find evidence of a range of behaviors in the high frequency regime of AGN PSDs estimated with multiple methods and in multiple data sets.  These findings imply that the DRW is not a universal model of AGN variability, but we do not necessarily understand what model is a more useful characterization.

Should a procedure for rehabilitating light curves for faint K2 targets become feasible, Figure~\ref{fig:fig16} provides key guidance with regard to the location and potential light curve quality of all the SDSS quasars that were targeted by our team.  Clearly the targets are not uniformly distributed just as the channels are not all of uniform photometric quality.  AGN appearing in channels outlined in cyan (light) blue may be the best choices for further investigation, while those those AGN appearing in channels outlined in purple and magenta, may be beyond repair.

\section{Summary}
\label{sec:summary}
{\em Kepler}/K2 has the potential to enable large-sample science programs for AGN, stars, and extra-galactic time domain phenomena. These large-sample programs are currently devastated by systematics that persist even after pointing errors known as arc-drift are corrected. 
The systematics may be similar to trends observed in the first phase of Kepler such as rolling band, Moir\'{e} noise, faint blended background sources, and changes in point spread function (PSF).
For variability studies requiring the full light curve PSD, there is no way to measure the performance trade-off of removing instrumental trends and scattering the astrophysical signal. Cotrending basis vectors (or PCA components), which were popular for detrending {\em Kepler} light curves, suffer this particular failing because they do not orthogonally account for magnitude or spatial information which we have shown to contribute significantly to the shape of the systematics. Careful error mitigation is even more critical to enable science for K2 programs at magnitudes and light curve cadence that are unmatched by other telescopes.     

We note two particularly useful characteristics of the systematics for readers interested in rehabilitating these data. First, there is a trade-off between arc-drift and the instrumental trends that we wish to remove.  Arc-drift is easier to remove and in some cases not present in larger aperture definitions.  However, the inclusion of more pixels in the aperture leads to stronger, more complex instrumental trends. 
Second, there appear to be some modules that are significantly less affected by systematics according to empty pixels as shown in Figure~\ref{fig:fig11}. Modules 14, 15, 19, and 20 may provide the best data quality, whereas modules 6, 11, and 16 exemplify the worst. This module quality, and by extension data quality, is independent of campaign or telescope orientation.   

The mirroring between Campaigns 8 and 16, in Figure~\ref{fig:fig4}, suggests that the telescope-Sun orientation 
may be responsible for an increasing or decreasing trend in some modules. The direction of the trend depends on whether the spacecraft was forward-facing or rear-facing, but the slope and amplitude of systematic features are consistently larger in poorer quality channels regardless of the spacecraft orientation.

An ideal reprocessing pipeline would consider campaign, module/channel number, magnitude,
 similar magnitude neighbors, aperture optimization, local background subtraction, arc-drift correction, and distance from the boresight. The {\tt lightkurve} package provides most of the functionality to get started.
 
The {\em Kepler}/K2 observations are a powerful legacy dataset for investigations of variable objects. The statistical criteria for doing AGN variability studies are rather different from the requirements for exoplanet discovery, for which the experiment was designed, thus some of the problems we identify above are particularly challenging. 
More precise characterization and amelioration of these systematic effects 
are beyond the scope of this work and, we believe, requires in-depth investigation by teams experienced in optical instrumentation.
Nevertheless, we anticipate that the effort required to clean {\em Kepler}/K2 light curves of the systematics we describe above will pay worthwhile dividends not only for AGN science, but also expand the parameter space of exoplanet detections that lie at even lower signal to noise than has been achieved to date.

\acknowledgements

We acknowledge support from NASA grant NNX17AF18G and NASA K2 guest observer grant 80NSSC19K0249.  This paper includes data collected by the K2 mission. Funding for the K2 mission is provided by the NASA Science Mission directorate.  M.S.V. acknowledges support from the Ambrose Mondell Foundation during sabbatical leave at the Institute for Advanced Study.  We thank Michael Gully-Santiago, Nicholas Saunders, Gert Barentsen, Patricia Boyd, and Thomas Barclay for their contributions to the discussion about this work.

\bibliographystyle{./aasjournal}
\bibliography{./ms.bib}

\begin{thebibliography}{}
\expandafter\ifx\csname natexlab\endcsname\relax\def\natexlab#1{#1}\fi
\providecommand{\url}[1]{\href{#1}{#1}}
\providecommand{\dodoi}[1]{doi:~\href{http://doi.org/#1}{\nolinkurl{#1}}}
\providecommand{\doeprint}[1]{\href{http://ascl.net/#1}{\nolinkurl{http://ascl.net/#1}}}
\providecommand{\doarXiv}[1]{\href{https://arxiv.org/abs/#1}{\nolinkurl{https://arxiv.org/abs/#1}}}

\bibitem[{{Aigrain} {et~al.}(2015){Aigrain}, {Hodgkin}, {Irwin}, {Lewis}, \&
  {Roberts}}]{Aigrain2015}
{Aigrain}, S., {Hodgkin}, S.~T., {Irwin}, M.~J., {Lewis}, J.~R., \& {Roberts},
  S.~J. 2015, \mnras, 447, 2880, \dodoi{10.1093/mnras/stu2638}

\bibitem[{{Annis} {et~al.}(2014){Annis}, {Soares-Santos}, {Strauss}, {Becker},
  {Dodelson}, {Fan}, {Gunn}, {Hao}, {Ivezi{\'c}}, {Jester}, {Jiang},
  {Johnston}, {Kubo}, {Lampeitl}, {Lin}, {Lupton}, {Miknaitis}, {Seo}, {Simet},
  \& {Yanny}}]{annis+14}
{Annis}, J., {Soares-Santos}, M., {Strauss}, M.~A., {et~al.} 2014, \apj, 794,
  120, \dodoi{10.1088/0004-637X/794/2/120}

\bibitem[{{Aranzana} {et~al.}(2018){Aranzana}, {K{\"o}rding}, {Uttley},
  {Scaringi}, \& {Bloemen}}]{Aranzana2018}
{Aranzana}, E., {K{\"o}rding}, E., {Uttley}, P., {Scaringi}, S., \& {Bloemen},
  S. 2018, \mnras, 476, 2501, \dodoi{10.1093/mnras/sty413}

\bibitem[{{Armstrong} {et~al.}(2015){Armstrong}, {Kirk}, {Lam}, {McCormac},
  {Walker}, {Brown}, {Osborn}, {Pollacco}, \& {Spake}}]{Armstrong2015}
{Armstrong}, D.~J., {Kirk}, J., {Lam}, K.~W.~F., {et~al.} 2015, \aap, 579, A19,
  \dodoi{10.1051/0004-6361/201525889}

\bibitem[{{Bellm} {et~al.}(2019){Bellm}, {Kulkarni}, {Graham}, {Dekany},
  {Smith}, {Riddle}, {Masci}, {Helou}, {Prince}, {Adams}, {Barbarino},
  {Barlow}, {Bauer}, {Beck}, {Belicki}, {Biswas}, {Blagorodnova}, {Bodewits},
  {Bolin}, {Brinnel}, {Brooke}, {Bue}, {Bulla}, {Burruss}, {Cenko}, {Chang},
  {Connolly}, {Coughlin}, {Cromer}, {Cunningham}, {De}, {Delacroix}, {Desai},
  {Duev}, {Eadie}, {Farnham}, {Feeney}, {Feindt}, {Flynn}, {Franckowiak},
  {Frederick}, {Fremling}, {Gal-Yam}, {Gezari}, {Giomi}, {Goldstein},
  {Golkhou}, {Goobar}, {Groom}, {Hacopians}, {Hale}, {Henning}, {Ho}, {Hover},
  {Howell}, {Hung}, {Huppenkothen}, {Imel}, {Ip}, {Ivezi{\'c}}, {Jackson},
  {Jones}, {Juric}, {Kasliwal}, {Kaspi}, {Kaye}, {Kelley}, {Kowalski},
  {Kramer}, {Kupfer}, {Landry}, {Laher}, {Lee}, {Lin}, {Lin}, {Lunnan},
  {Giomi}, {Mahabal}, {Mao}, {Miller}, {Monkewitz}, {Murphy}, {Ngeow},
  {Nordin}, {Nugent}, {Ofek}, {Patterson}, {Penprase}, {Porter}, {Rauch},
  {Rebbapragada}, {Reiley}, {Rigault}, {Rodriguez}, {van Roestel}, {Rusholme},
  {van Santen}, {Schulze}, {Shupe}, {Singer}, {Soumagnac}, {Stein}, {Surace},
  {Sollerman}, {Szkody}, {Taddia}, {Terek}, {Van Sistine}, {van Velzen},
  {Vestrand}, {Walters}, {Ward}, {Ye}, {Yu}, {Yan}, \& {Zolkower}}]{Bellm2019}
{Bellm}, E.~C., {Kulkarni}, S.~R., {Graham}, M.~J., {et~al.} 2019, \pasp, 131,
  018002, \dodoi{10.1088/1538-3873/aaecbe}

\bibitem[{{Bovy} {et~al.}(2011){Bovy}, {Hennawi}, {Hogg}, {Myers},
  {Kirkpatrick}, {Schlegel}, {Ross}, {Sheldon}, {McGreer}, {Schneider}, \&
  {Weaver}}]{bhh+11}
{Bovy}, J., {Hennawi}, J.~F., {Hogg}, D.~W., {et~al.} 2011, \apj, 729, 141,
  \dodoi{10.1088/0004-637X/729/2/141}

\bibitem[{Bryson {et~al.}(2010)Bryson, Tenenbaum, Jenkins, Chandrasekaran,
  Klaus, Caldwell, Gilliland, Haas, Dotson, Koch, \& Borucki}]{Bryson2010}
Bryson, S.~T., Tenenbaum, P., Jenkins, J.~M., {et~al.} 2010, The Astrophysical
  Journal, 713, L97, \dodoi{10.1088/2041-8205/713/2/l97}

\bibitem[{{Caplar} {et~al.}(2017){Caplar}, {Lilly}, \&
  {Trakhtenbrot}}]{Caplar2017}
{Caplar}, N., {Lilly}, S.~J., \& {Trakhtenbrot}, B. 2017, \apj, 834, 111,
  \dodoi{10.3847/1538-4357/834/2/111}

\bibitem[{{Chen} \& {Wang}(2015)}]{Chen2015}
{Chen}, X.-Y., \& {Wang}, J.-X. 2015, \apj, 805, 80,
  \dodoi{10.1088/0004-637X/805/1/80}

\bibitem[{Cleve {et~al.}(2016)Cleve, Howell, Smith, Clarke, Thompson, Bryson,
  Lund, Handberg, \& Chaplin}]{Cleve2016}
Cleve, J. E.~V., Howell, S.~B., Smith, J.~C., {et~al.} 2016, Publications of
  the Astronomical Society of the Pacific, 128, 075002,
  \dodoi{10.1088/1538-3873/128/965/075002}

\bibitem[{{Crossfield} {et~al.}(2015){Crossfield}, {Petigura}, {Schlieder},
  {Howard}, {Fulton}, {Aller}, {Ciardi}, {L{\'e}pine}, {Barclay}, {de Pater},
  {de Kleer}, {Quintana}, {Christiansen}, {Schlafly}, {Kaltenegger}, {Crepp},
  {Henning}, {Obermeier}, {Deacon}, {Weiss}, {Isaacson}, {Hansen}, {Liu},
  {Greene}, {Howell}, {Barman}, \& {Mordasini}}]{Crossfield2015}
{Crossfield}, I. J.~M., {Petigura}, E., {Schlieder}, J.~E., {et~al.} 2015,
  \apj, 804, 10, \dodoi{10.1088/0004-637X/804/1/10}

\bibitem[{Deming {et~al.}(2015)Deming, Knutson, Kammer, Fulton, Ingalls, Carey,
  Burrows, Fortney, Todorov, Agol, Cowan, Desert, Fraine, Langton, Morley, \&
  Showman}]{Deming2015}
Deming, D., Knutson, H., Kammer, J., {et~al.} 2015, The Astrophysical Journal,
  805, 132, \dodoi{10.1088/0004-637x/805/2/132}

\bibitem[{{Dobrotka} {et~al.}(2017){Dobrotka}, {Antonuccio-Delogu}, \&
  {Baj{\v{c}}i{\v{c}}{\'a}kov{\'a}}}]{Dobrotka2017}
{Dobrotka}, A., {Antonuccio-Delogu}, V., \& {Baj{\v{c}}i{\v{c}}{\'a}kov{\'a}},
  I. 2017, \mnras, 470, 2439, \dodoi{10.1093/mnras/stx961}

\bibitem[{{Dobrotka} {et~al.}(2019){Dobrotka}, {Bez{\'a}k}, {Revalski}, \&
  {Str{\'e}my}}]{Dobrotka2019}
{Dobrotka}, A., {Bez{\'a}k}, P., {Revalski}, M., \& {Str{\'e}my}, M. 2019,
  \mnras, 483, 38, \dodoi{10.1093/mnras/sty3074}

\bibitem[{{Edelson} \& {Malkan}(2012)}]{Edelson2012}
{Edelson}, R., \& {Malkan}, M. 2012, \apj, 751, 52,
  \dodoi{10.1088/0004-637X/751/1/52}

\bibitem[{{Edelson} {et~al.}(2014){Edelson}, {Vaughan}, {Malkan}, {Kelly},
  {Smith}, {Boyd}, \& {Mushotzky}}]{Edelson2014}
{Edelson}, R., {Vaughan}, S., {Malkan}, M., {et~al.} 2014, \apj, 795, 2,
  \dodoi{10.1088/0004-637X/795/1/2}

\bibitem[{{Foreman-Mackey} {et~al.}(2017){Foreman-Mackey}, {Agol},
  {Ambikasaran}, \& {Angus}}]{Celerite}
{Foreman-Mackey}, D., {Agol}, E., {Ambikasaran}, S., \& {Angus}, R. 2017, \aj,
  154, 220, \dodoi{10.3847/1538-3881/aa9332}

\bibitem[{{Foreman-Mackey} {et~al.}(2015){Foreman-Mackey}, {Montet}, {Hogg},
  {Morton}, {Wang}, \& {Sch{\"o}lkopf}}]{Foreman-Mackey2015}
{Foreman-Mackey}, D., {Montet}, B.~T., {Hogg}, D.~W., {et~al.} 2015, \apj, 806,
  215, \dodoi{10.1088/0004-637X/806/2/215}

\bibitem[{{Giustini} \& {Proga}(2019)}]{GP19}
{Giustini}, M., \& {Proga}, D. 2019, \aap, 630, A94,
  \dodoi{10.1051/0004-6361/201833810}

\bibitem[{{Howell} {et~al.}(2014){Howell}, {Sobeck}, {Haas}, {Still},
  {Barclay}, {Mullally}, {Troeltzsch}, {Aigrain}, {Bryson}, {Caldwell},
  {Chaplin}, {Cochran}, {Huber}, {Marcy}, {Miglio}, {Najita}, {Smith},
  {Twicken}, \& {Fortney}}]{Howell+14}
{Howell}, S.~B., {Sobeck}, C., {Haas}, M., {et~al.} 2014, \pasp, 126, 398,
  \dodoi{10.1086/676406}

\bibitem[{{Huang} {et~al.}(2015){Huang}, {Penev}, {Hartman}, {Bakos}, {Bhatti},
  {Domsa}, \& {de Val-Borro}}]{Huang2015}
{Huang}, C.~X., {Penev}, K., {Hartman}, J.~D., {et~al.} 2015, \mnras, 454,
  4159, \dodoi{10.1093/mnras/stv2257}

\bibitem[{{Ivezi{\'c}} {et~al.}(2008){Ivezi{\'c}}, {Axelrod}, {Becker},
  {Becla}, {Borne}, {Burke}, {Claver}, {Cook}, {Connolly}, {Gilmore}, {Jones},
  {Juri{\'c}}, {Kahn}, {Lim}, {Lupton}, {Monet}, {Pinto}, {Sesar}, {Stubbs}, \&
  {Tyson}}]{Ivezic2008}
{Ivezi{\'c}}, {\v{Z}}., {Axelrod}, T., {Becker}, A.~C., {et~al.} 2008, in
  American Institute of Physics Conference Series, ed. C.~A.~L. {Bailer-Jones},
  Vol. 1082, 359--365, \dodoi{10.1063/1.3059076}

\bibitem[{{Kasliwal} {et~al.}(2017){Kasliwal}, {Vogeley}, \&
  {Richards}}]{Vish2017}
{Kasliwal}, V.~P., {Vogeley}, M.~S., \& {Richards}, G.~T. 2017, \mnras, 470,
  3027, \dodoi{10.1093/mnras/stx1420}

\bibitem[{{Kasliwal} {et~al.}(2015){Kasliwal}, {Vogeley}, {Richards},
  {Williams}, \& {Carini}}]{Vish2015}
{Kasliwal}, V.~P., {Vogeley}, M.~S., {Richards}, G.~T., {Williams}, J., \&
  {Carini}, M.~T. 2015, \mnras, 453, 2075, \dodoi{10.1093/mnras/stv1797}

\bibitem[{{Kinemuchi} {et~al.}(2012){Kinemuchi}, {Barclay}, {Fanelli},
  {Pepper}, {Still}, \& {Howell}}]{Kinemuchi2012}
{Kinemuchi}, K., {Barclay}, T., {Fanelli}, M., {et~al.} 2012, \pasp, 124, 963,
  \dodoi{10.1086/667603}

\bibitem[{Libralato {et~al.}(2015)Libralato, Bedin, Nardiello, \&
  Piotto}]{Libralato2015}
Libralato, M., Bedin, L.~R., Nardiello, D., \& Piotto, G. 2015, Monthly Notices
  of the Royal Astronomical Society, 456, 1137, \dodoi{10.1093/mnras/stv2628}

\bibitem[{{Lightkurve Collaboration} {et~al.}(2018){Lightkurve Collaboration},
  {Cardoso}, {Hedges}, {Gully-Santiago}, {Saunders}, {Cody}, {Barclay}, {Hall},
  {Sagear}, {Turtelboom}, {Zhang}, {Tzanidakis}, {Mighell}, {Coughlin}, {Bell},
  {Berta-Thompson}, {Williams}, {Dotson}, \& {Barentsen}}]{lightkurve2018}
{Lightkurve Collaboration}, {Cardoso}, J.~V.~d.~M., {Hedges}, C., {et~al.}
  2018, {Lightkurve: Kepler and TESS time series analysis in Python},
  Astrophysics Source Code Library.
\newblock \doeprint{1812.013}

\bibitem[{Luger {et~al.}(2016)Luger, Agol, Kruse, Barnes, Becker,
  Foreman-Mackey, \& Deming}]{Luger2016}
Luger, R., Agol, E., Kruse, E., {et~al.} 2016, The Astronomical Journal, 152,
  100, \dodoi{10.3847/0004-6256/152/4/100}

\bibitem[{{Luger} {et~al.}(2018){Luger}, {Kruse}, {Foreman-Mackey}, {Agol}, \&
  {Saunders}}]{Lugar2018}
{Luger}, R., {Kruse}, E., {Foreman-Mackey}, D., {Agol}, E., \& {Saunders}, N.
  2018, \aj, 156, 99, \dodoi{10.3847/1538-3881/aad230}

\bibitem[{{Lund} {et~al.}(2015){Lund}, {Handberg}, {Davies}, {Chaplin}, \&
  {Jones}}]{Lund2015}
{Lund}, M.~N., {Handberg}, R., {Davies}, G.~R., {Chaplin}, W.~J., \& {Jones},
  C.~D. 2015, \apj, 806, 30, \dodoi{10.1088/0004-637X/806/1/30}

\bibitem[{McCalmont {et~al.}(2015)McCalmont, Larson, Peterson, \&
  Ross}]{McCalmont2015}
McCalmont, K.~M., Larson, K.~A., Peterson, C.~A., \& Ross, S.~E. 2015

\bibitem[{Moreno {et~al.}(2019)Moreno, Vogeley, Richards, \& Yu}]{Moreno2019}
Moreno, J., Vogeley, M.~S., Richards, G.~T., \& Yu, W. 2019, Publications of
  the Astronomical Society of the Pacific, 131, 063001,
  \dodoi{10.1088/1538-3873/ab1597}

\bibitem[{{Mushotzky} {et~al.}(2011){Mushotzky}, {Edelson}, {Baumgartner}, \&
  {Gandhi}}]{Mushotzky2011}
{Mushotzky}, R.~F., {Edelson}, R., {Baumgartner}, W., \& {Gandhi}, P. 2011,
  \apj, 743, L12, \dodoi{10.1088/2041-8205/743/1/L12}

\bibitem[{{O'Brien} {et~al.}(2018){O'Brien}, {Moreno}, {Richards}, \&
  {Vogeley}}]{jack2018}
{O'Brien}, J.~T., {Moreno}, J., {Richards}, G.~T., \& {Vogeley}, M.~S. 2018,
  Research Notes of the American Astronomical Society, 2, 127,
  \dodoi{10.3847/2515-5172/aad331}

\bibitem[{{P{\^a}ris} {et~al.}(2017){P{\^a}ris}, {Petitjean}, {Ross}, {Myers},
  {Aubourg}, {Streblyanska}, {Bailey}, {Armengaud}, {Palanque-Delabrouille},
  {Y{\`e}che}, {Hamann}, {Strauss}, {Albareti}, {Bovy}, {Bizyaev}, {Niel
  Brandt}, {Brusa}, {Buchner}, {Comparat}, {Croft}, {Dwelly}, {Fan},
  {Font-Ribera}, {Ge}, {Georgakakis}, {Hall}, {Jiang}, {Kinemuchi},
  {Malanushenko}, {Malanushenko}, {McMahon}, {Menzel}, {Merloni}, {Nandra},
  {Noterdaeme}, {Oravetz}, {Pan}, {Pieri}, {Prada}, {Salvato}, {Schlegel},
  {Schneider}, {Simmons}, {Viel}, {Weinberg}, \& {Zhu}}]{DR12Q}
{P{\^a}ris}, I., {Petitjean}, P., {Ross}, N.~P., {et~al.} 2017, \aap, 597, A79,
  \dodoi{10.1051/0004-6361/201527999}

\bibitem[{{Peters} {et~al.}(2015){Peters}, {Richards}, {Myers}, {Strauss},
  {Schmidt}, {Ivezi{\'c}}, {Ross}, {MacLeod}, \& {Riegel}}]{Peters15}
{Peters}, C.~M., {Richards}, G.~T., {Myers}, A.~D., {et~al.} 2015, \apj, 811,
  95, \dodoi{10.1088/0004-637X/811/2/95}

\bibitem[{{Revalski} {et~al.}(2014){Revalski}, {Nowak}, {Wiita}, {Wehrle}, \&
  {Unwin}}]{Revalski2014}
{Revalski}, M., {Nowak}, D., {Wiita}, P.~J., {Wehrle}, A.~E., \& {Unwin}, S.~C.
  2014, \apj, 785, 60, \dodoi{10.1088/0004-637X/785/1/60}

\bibitem[{{Richards} {et~al.}(2004){Richards}, {Nichol}, {Gray}, {Brunner},
  {Lupton}, {Vanden Berk}, {Chong}, {Weinstein}, {Schneider}, {Anderson},
  {Munn}, {Harris}, {Strauss}, {Fan}, {Gunn}, {Ivezi{\'c}}, {York},
  {Brinkmann}, \& {Moore}}]{rng+04}
{Richards}, G.~T., {Nichol}, R.~C., {Gray}, A.~G., {et~al.} 2004, \apjs, 155,
  257, \dodoi{10.1086/425356}

\bibitem[{{Richards} {et~al.}(2015){Richards}, {Myers}, {Peters}, {Krawczyk},
  {Chase}, {Ross}, {Fan}, {Jiang}, {Lacy}, {McGreer}, {Trump}, \&
  {Riegel}}]{Richards15}
{Richards}, G.~T., {Myers}, A.~D., {Peters}, C.~M., {et~al.} 2015, \apjs, 219,
  39, \dodoi{10.1088/0067-0049/219/2/39}

\bibitem[{{Saunders} {et~al.}(2019){Saunders}, {Luger}, \&
  {Barnes}}]{Saunders2019}
{Saunders}, N., {Luger}, R., \& {Barnes}, R. 2019, \aj, 157, 197,
  \dodoi{10.3847/1538-3881/ab12e4}

\bibitem[{{Schneider} {et~al.}(2010){Schneider}, {Richards}, {Hall}, {Strauss},
  {Anderson}, {Boroson}, {Ross}, {Shen}, {Brandt}, {Fan}, {Inada}, {Jester},
  {Knapp}, {Krawczyk}, {Thakar}, {Vanden Berk}, {Voges}, {Yanny}, {York},
  {Bahcall}, {Bizyaev}, {Blanton}, {Brewington}, {Brinkmann}, {Eisenstein},
  {Frieman}, {Fukugita}, {Gray}, {Gunn}, {Hibon}, {Ivezi{\'c}}, {Kent}, {Kron},
  {Lee}, {Lupton}, {Malanushenko}, {Malanushenko}, {Oravetz}, {Pan}, {Pier},
  {Price}, {Saxe}, {Schlegel}, {Simmons}, {Snedden}, {SubbaRao}, {Szalay}, \&
  {Weinberg}}]{srh+10}
{Schneider}, D.~P., {Richards}, G.~T., {Hall}, P.~B., {et~al.} 2010, \aj, 139,
  2360, \dodoi{10.1088/0004-6256/139/6/2360}

\bibitem[{{Shaya} {et~al.}(2015){Shaya}, {Olling}, \& {Mushotzky}}]{EShaya2015}
{Shaya}, E.~J., {Olling}, R., \& {Mushotzky}, R. 2015, \aj, 150, 188,
  \dodoi{10.1088/0004-6256/150/6/188}

\bibitem[{{Simm} {et~al.}(2016){Simm}, {Salvato}, {Saglia}, {Ponti},
  {Lanzuisi}, {Trakhtenbrot}, {Nandra}, \& {Bender}}]{Simm2016}
{Simm}, T., {Salvato}, M., {Saglia}, R., {et~al.} 2016, \aap, 585, A129,
  \dodoi{10.1051/0004-6361/201527353}

\bibitem[{{Smith} {et~al.}(2012){Smith}, {Stumpe}, {Van Cleve}, {Jenkins},
  {Barclay}, {Fanelli}, {Girouard}, {Kolodziejczak}, {McCauliff}, {Morris}, \&
  {Twicken}}]{Smith2012}
{Smith}, J.~C., {Stumpe}, M.~C., {Van Cleve}, J.~E., {et~al.} 2012, \pasp, 124,
  1000, \dodoi{10.1086/667697}

\bibitem[{{Smith} {et~al.}(2018{\natexlab{a}}){Smith}, {Mushotzky}, {Boyd},
  {Malkan}, {Howell}, \& {Gelino}}]{Smith2018}
{Smith}, K.~L., {Mushotzky}, R.~F., {Boyd}, P.~T., {et~al.} 2018{\natexlab{a}},
  \apj, 857, 141, \dodoi{10.3847/1538-4357/aab88d}

\bibitem[{{Smith} {et~al.}(2018{\natexlab{b}}){Smith}, {Mushotzky}, {Boyd}, \&
  {Wagoner}}]{Krista2018b}
{Smith}, K.~L., {Mushotzky}, R.~F., {Boyd}, P.~T., \& {Wagoner}, R.~V.
  2018{\natexlab{b}}, \apj, 860, L10, \dodoi{10.3847/2041-8213/aac88c}

\bibitem[{{Smith} {et~al.}(2015){Smith}, {Boyd}, {Mushotzky}, {Gehrels},
  {Edelson}, {Howell}, {Gelino}, {Brown}, \& {Young}}]{KSmith2015}
{Smith}, K.~L., {Boyd}, P.~T., {Mushotzky}, R.~F., {et~al.} 2015, \aj, 150,
  126, \dodoi{10.1088/0004-6256/150/4/126}

\bibitem[{{Still} \& {Barclay}(2012)}]{Still2012}
{Still}, M., \& {Barclay}, T. 2012, {PyKE: Reduction and analysis of Kepler
  Simple Aperture Photometry data}, Astrophysics Source Code Library.
\newblock \doeprint{1208.004}

\bibitem[{{Stumpe} {et~al.}(2012){Stumpe}, {Smith}, {Van Cleve}, {Twicken},
  {Barclay}, {Fanelli}, {Girouard}, {Jenkins}, {Kolodziejczak}, {McCauliff}, \&
  {Morris}}]{Stumpe2012part1}
{Stumpe}, M.~C., {Smith}, J.~C., {Van Cleve}, J.~E., {et~al.} 2012, \pasp, 124,
  985, \dodoi{10.1086/667698}

\bibitem[{{Tamuz} {et~al.}(2005){Tamuz}, {Mazeh}, \& {Zucker}}]{Tamuz2005}
{Tamuz}, O., {Mazeh}, T., \& {Zucker}, S. 2005, \mnras, 356, 1466,
  \dodoi{10.1111/j.1365-2966.2004.08585.x}

\bibitem[{{Uttley} {et~al.}(2002){Uttley}, {McHardy}, \&
  {Papadakis}}]{UttleyPSD}
{Uttley}, P., {McHardy}, I.~M., \& {Papadakis}, I.~E. 2002, \mnras, 332, 231,
  \dodoi{10.1046/j.1365-8711.2002.05298.x}

\bibitem[{{Van Cleve} \& {Bryson}(2017)}]{Cleve2017}
{Van Cleve}, J.~E., \& {Bryson}, S.~T. 2017, {K2 Handbook}, Tech. rep.

\bibitem[{{Van Cleve} \& {Caldwell}(2016)}]{KeplerHandbook}
{Van Cleve}, J.~E., \& {Caldwell}, D.~A. 2016, {Kepler Instrument Handbook},
  Tech. rep.

\bibitem[{{Vanderburg} \& {Johnson}(2014)}]{VJ2014}
{Vanderburg}, A., \& {Johnson}, J.~A. 2014, \pasp, 126, 948,
  \dodoi{10.1086/678764}

\bibitem[{{V{\'e}ron-Cetty} \& {V{\'e}ron}(2010)}]{VCV2010}
{V{\'e}ron-Cetty}, M.-P., \& {V{\'e}ron}, P. 2010, \aap, 518, A10,
  \dodoi{10.1051/0004-6361/201014188}

\bibitem[{{Wehrle} {et~al.}(2013){Wehrle}, {Wiita}, {Unwin}, {Di Lorenzo},
  {Revalski}, {Silano}, \& {Sprague}}]{Wehrle2013}
{Wehrle}, A.~E., {Wiita}, P.~J., {Unwin}, S.~C., {et~al.} 2013, \apj, 773, 89,
  \dodoi{10.1088/0004-637X/773/2/89}

\bibitem[{{York} {et~al.}(2000){York}, {Adelman}, {Anderson}, {Anderson},
  {Annis}, {Bahcall}, {Bakken}, {Barkhouser}, {Bastian}, {Berman}, {Boroski},
  {Bracker}, {Briegel}, {Briggs}, {Brinkmann}, {Brunner}, {Burles}, {Carey},
  {Carr}, {Castander}, {Chen}, {Colestock}, {Connolly}, {Crocker}, {Csabai},
  {Czarapata}, {Davis}, {Doi}, {Dombeck}, {Eisenstein}, {Ellman}, {Elms},
  {Evans}, {Fan}, {Federwitz}, {Fiscelli}, {Friedman}, {Frieman}, {Fukugita},
  {Gillespie}, {Gunn}, {Gurbani}, {de Haas}, {Haldeman}, {Harris}, {Hayes},
  {Heckman}, {Hennessy}, {Hindsley}, {Holm}, {Holmgren}, {Huang}, {Hull},
  {Husby}, {Ichikawa}, {Ichikawa}, {Ivezi{\'c}}, {Kent}, {Kim}, {Kinney},
  {Klaene}, {Kleinman}, {Kleinman}, {Knapp}, {Korienek}, {Kron}, {Kunszt},
  {Lamb}, {Lee}, {Leger}, {Limmongkol}, {Lindenmeyer}, {Long}, {Loomis},
  {Loveday}, {Lucinio}, {Lupton}, {MacKinnon}, {Mannery}, {Mantsch}, {Margon},
  {McGehee}, {McKay}, {Meiksin}, {Merelli}, {Monet}, {Munn}, {Narayanan},
  {Nash}, {Neilsen}, {Neswold}, {Newberg}, {Nichol}, {Nicinski}, {Nonino},
  {Okada}, {Okamura}, {Ostriker}, {Owen}, {Pauls}, {Peoples}, {Peterson},
  {Petravick}, {Pier}, {Pope}, {Pordes}, {Prosapio}, {Rechenmacher}, {Quinn},
  {Richards}, {Richmond}, {Rivetta}, {Rockosi}, {Ruthmansdorfer}, {Sand ford},
  {Schlegel}, {Schneider}, {Sekiguchi}, {Sergey}, {Shimasaku}, {Siegmund},
  {Smee}, {Smith}, {Snedden}, {Stone}, {Stoughton}, {Strauss}, {Stubbs},
  {SubbaRao}, {Szalay}, {Szapudi}, {Szokoly}, {Thakar}, {Tremonti}, {Tucker},
  {Uomoto}, {Vanden Berk}, {Vogeley}, {Waddell}, {Wang}, {Watanabe},
  {Weinberg}, {Yanny}, {Yasuda}, \& {SDSS Collaboration}}]{York2000}
{York}, D.~G., {Adelman}, J., {Anderson}, John~E., J., {et~al.} 2000, \aj, 120,
  1579, \dodoi{10.1086/301513}

\bibitem[{{Zinn} {et~al.}(2017){Zinn}, {Kochanek}, {Koz{\l}owski}, {Udalski},
  {Szyma{\'n}ski}, {Soszy{\'n}ski}, {Wyrzykowski}, {Ulaczyk}, {Poleski},
  {Pietrukowicz}, {Skowron}, {Mr{\'o}z}, \& {Pawlak}}]{Zinn2017}
{Zinn}, J.~C., {Kochanek}, C.~S., {Koz{\l}owski}, S., {et~al.} 2017, \mnras,
  468, 2189, \dodoi{10.1093/mnras/stx586}

\end{thebibliography}

\end{document}